\documentclass[preprint]{elsarticle}
\usepackage{graphicx,url,hyperref,microtype,subcaption,amsmath,graphbox}
\usepackage{latexsym,courier,upgreek,xcolor,rotating,soul}
\usepackage{algorithmic}
\usepackage[displaymath,mathlines]{lineno}
\usepackage[margin=2cm]{geometry}
\hypersetup{colorlinks,citecolor=blue,linkcolor=red,urlcolor=green}
\modulolinenumbers[5]
\biboptions{numbers,sort&compress}

\makeatletter
\def\ps@pprintTitle{%
 \let\@oddhead\@empty
 \let\@evenhead\@empty
 \def\@oddfoot{}%
 \let\@evenfoot\@oddfoot}
\makeatother

\begin{document}

\begin{frontmatter}
    \title{Mixed-wet percolation on a dual square lattice}
    \author[add1]{Jnana Ranjan Das}
    \ead{d.jnana@iitg.ac.in}

    \author[add2]{Santanu Sinha}
    \ead{santanu.sinha@ntnu.no}

    \author[add2,add1]{Alex Hansen}
    \ead{alex.hansen@ntnu.no}

    \author[add1]{Sitangshu B. Santra}
    \ead{santra@iitg.ac.in}

    \address[add1]{Department of Physics, Indian Institute of Technology Guwahati, Guwahati--781039 Assam, India}
    \address[add2]{PoreLab, Department of Physics, Norwegian University of Science and Technology, NO--7491 Trondheim, Norway}

    \begin{abstract}
        We present a percolation model that is inspired by recent works on immiscible two-phase flow in a mixed-wet porous medium made of a mixture of grains with two different wettability properties. The percolation model is constructed on a dual lattice where the sites on the primal lattice represent the grains of the porous medium, and the bonds on the dual lattice represent the pores in between the grains. The bonds on the dual lattice are occupied based on the two adjacent sites on the primal lattice, which represent the pores where the capillary forces average to zero. The spanning cluster of the bonds, therefore, represents the flow network through which the two immiscible fluids can flow without facing any capillary barrier. It turns out to be a percolation transition of the perimeters of a site percolation problem. We study the geometrical properties at the criticality of the perimeter system numerically. A scaling theory is developed for these properties, and their scaling relations with the respective density parameters are studied. We also verified their finite-size scaling relations. Though the site clusters and their perimeters look very different compared to ordinary percolation, the singular behaviour of the associated geometrical properties remains unchanged. The critical exponents are found to be those of the ordinary percolation.
    \end{abstract}
\end{frontmatter}

\section{Introduction}
\label{intro}
Percolation theory was first recognized as a critical phenomenon by Broadbent and Hammersley in 1957 \cite{broadbent1957percolation}.  Little did they know that their paper would give rise to literally thousands of papers in the ensuing years.  It remains the foremost example of a non-thermal critical phenomenon \cite{stauffer2018introduction,complexity}. The percolation problem in its primal form is easy to grasp.  However, the percolation problem comes in many disguises.  We may list some. Ambogaekar, Halperin and Langer demonstrated that the conductivity of strongly disordered conductors is related to a percolation threshold \cite{ambegaokar1971hopping}. A similar line of reasoning for measuring the permeability of porous rocks is behind the well-known Katz and Thompson method of mercury injection  \cite{katz1986quantitative}. Slow injection of one fluid into a porous medium already containing another fluid, and the two fluids are immiscible, is well modeled by the invasion percolation model \cite{wilkinson1983invasion}. A more recent version of the percolation problem that has generated considerable interest is explosive percolation \cite{achlioptas2009explosive,riordan2011explosive}.

\begin{figure}[t]
    \centerline{\hfill
        \includegraphics[width=0.3\textwidth,clip]{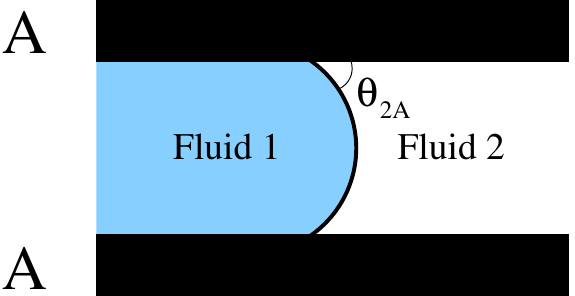}\hfill
        \includegraphics[width=0.3\textwidth,clip]{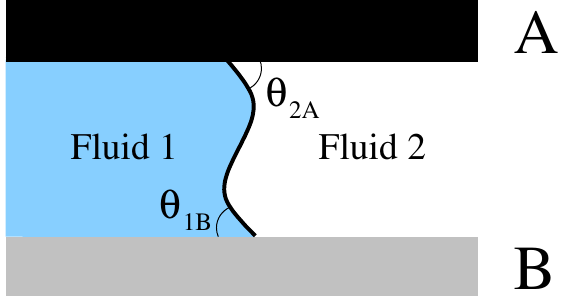}\hfill}
    \caption{\label{2PhF} Illustration of uniformly wet (left) and mixed wet (right) conditions inside a pore. On the left, the pore is in between two grains A with the same wettability type (black), which is less wetting to fluid 1 (blue) and more wetting to fluid 2 (white). The pore at the right side on the other hand is in between walls from two different grains A (black) and B (grey) with two opposite wettability types, where A is less wetting to fluid 1 and more wetting to fluid 2, whereas for B it is the opposite.}
\end{figure}

We present here a new percolation model that is motivated by the flow of immiscible fluids in a mixed-wet porous medium. Percolation models have widely been used in the past and present to characterise porous media flow, especially in the capillary-dominated regime where the flow is governed by the disorder in the pores \cite{sahimi2011flow, blunt2017multiphase, feder2022physics, hs17, ga19}. For example, the displacement of an immiscible fluid by another fluid inside a porous medium during capillary-dominated flow produces fingers due to capillary instabilities. These fingers are called capillary fingers, which were characterized by the invasion percolation model \cite{ltz88, lz89}. The invasion percolation model has also been used to find the relative permeability and saturation profile near the percolation critical point for a static two-phase flow system in the capillary-dominated regime \cite{w86}. The percolation model we present here is motivated by the experimental studies \cite{irannezhad2023fluid,irannezhad2023characteristics}, and the computational modeling \cite{fyhn2023effective} of two-phase flow in an engineered mixed-wet porous medium that is made of a random mixture of two types grains with opposite wetting properties with respect to the fluids flowing in the medium. Imagine a random mixture of two types of grains, say $A$ and $B$, and two immiscible fluids, say fluid $1$ and fluid $2$. Suppose the grain type $A$ is less wetting to fluid $1$ compared to fluid $2$, i.e., the wetting angle $\theta_{2A}< \pi/2$, see Fig. \ref{2PhF}. The grain type $B$, on the other hand, is more wetting to fluid $1$ compared to fluid $2$, i.e., $\theta_{1B}<\pi/2$.  Suppose now we have an interface between fluid $1$ and $2$ moving in a pore with $A$-type walls. The capillary force between the two fluids will then point towards fluid $1$ as fluid $2$ is more wetting than fluid $1$, see Fig. \ref{2PhF} (left).  If the walls are both of type $B$, the capillary force at the interface will point towards fluid $2$.  However, if one wall is of type $A$ and the other of type $B$, we have the situation shown in Fig. \ref{2PhF} (right). In this case, the capillary forces will be much weaker than the two previous cases as the contribution from each wall essentially cancels.             

Geistlinger et al.\ \cite{geistlinger2021new} did a micro-computed tomography ($\mu$-CT) study of a mixture of sand grains with contact angle either being $0^\circ$ or $100^\circ$ with respect to the fluid interfaces.  They then injected one of the fluids into the ensuing porous medium saturated with the other fluid.  They would recognize a structural percolation transition in the clusters of the defending fluid after the invasion process. Irannezahad et al.\ \cite{irannezhad2023fluid,irannezhad2023characteristics} studied the same problem numerically but not in a percolation context. Fyhn et al.\ \cite{fyhn2023effective} studied steady-state flow in mixed-wet porous media using a dynamic pore-network model \cite{sgv21}. By steady-state flow, we mean flow where the macroscopic averages of the various variables are constant or fluctuate around well-defined averages, and there are no macroscopic saturation gradients present \cite{erpelding2013history}.  Fyhn et al.\ \cite{fyhn2023effective} recognized that the steady-state mixed wet problem is a percolation problem in disguise.  

In 2009, Tallakstad et al.\ \cite{tallakstad2009steady,tallakstad2009steadyb} demonstrated that there is a flow regime under immiscible steady-state two-phase flow in porous media where the flow rate is proportional to the pressure gradient raised to a power close to two. This is due to interfaces held in place by capillary forces would be mobilized with increasing pressure drop. Sinha and Hansen \cite{sinha2012effective} found that if pressure drop rather than flow rate was the control variable, there typically would be a threshold pressure drop for flow to occur. This would be caused by the immobilized interfaces blocking the flow.  The presence of this threshold would make the determination of the exponent in the power law between flow rate and pressure drop difficult to determine as the threshold pressure drop would have to be determined simultaneously with the exponent.  This is where the mixed wet percolation problem enters.  With the saturation (mixing ratio) being in the correct range, connected paths constituting pores between two opposite types of walls $A$ and $B$ would appear, where there would essentially be no capillary forces. It is the structure of these paths between $A$ and $B$, which form the percolation problem we focus on here. We refer to this variant of the percolation problem as {\it mixed wet percolation\/}. 

In the following, we develop the mixed-wet percolation (MWP) model in Section \ref{model} to generate the network of zero capillary barrier paths for two-phase flow in a mixed-wet porous media, which is constituted of grains with opposite wettability properties. We model the porous media by ordinary site percolation (OSP) on a primal lattice. The network of mixed-wet pores in between the sites is then constituted of the links in its dual lattice between two adjacent occupied-unoccupied sites on the primal lattice. The network of the zero capillary barrier paths through which the two-phase flow is likely to start is found to be the perimeters of the cluster of either the occupied sites or the unoccupied sites, depending on the concentration of occupied (unoccupied) sites. In Section \ref{scaling}, we outline the scaling theory to study the critical behaviour of the geometrical quantities of the networks. We will present the numerical results in Section \ref{rd} and verify the scaling properties. We draw our conclusions in Section \ref{conclusion}.

\section{The mixed-wet percolation (MWP) Model}
\label{model}
\begin{figure}[t]
    \centerline{\hfill
        \includegraphics[width=0.15\linewidth,clip]{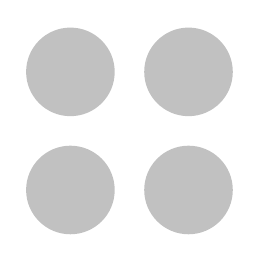}\qquad
        \includegraphics[width=0.15\linewidth,clip]{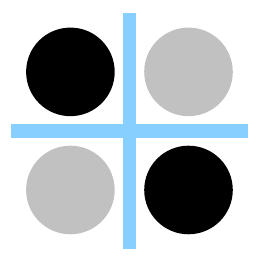}\qquad
        \includegraphics[width=0.15\linewidth,clip]{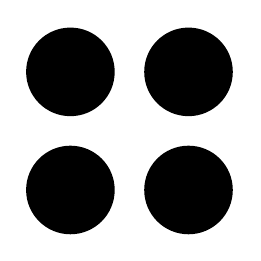}\hfill}
    \centerline{\hspace{5cm} (a)\hspace{3cm} (b)\hspace{2.7cm} (c)\hfill}
    \caption{\label{Fl2d} The black circles represent $A$-type grains and the grey circles represent $B$-type grains. In (a), all grains are $B$-type. Hence, there will be no flow, and no links on the dual lattice will be occupied. Similarly, in (c), all the grains are $A$-type. It corresponds to no flow, and no links are occupied. In (b), two $A$-type and two $B$-type grains are placed diagonally opposite such that they are always adjacently different, and fluid can flow through all the links between them. Hence, they are occupied}
\end{figure}

Since the square lattice is a self-dual lattice, it is easy to model mixed wet percolation involving a primal square lattice and its self-dual lattice in $2$d. The ordinary site percolation is implemented on the primal lattice. Starting from an empty primal lattice, the sites are occupied randomly with a probability $p$ and the rest of the sites remain unoccupied with a probability $(1-p)$. The occupied and unoccupied sites represent $A$ and $B$ type grains, respectively. Then, we consider the links on the dual lattice. If a link of the dual lattice is either between two adjacent occupied sites or between two adjacent unoccupied sites on the primal lattice, no bond is placed at that link on the dual lattice. If, however, the two adjacent lattice sites on the primal lattice are of different types, i.e. occupied-unoccupied or vice versa, we place a bond on the link of the dual lattice with unit probability. It is demonstrated with a $2\times 2$ square lattice and two different grains in Fig. \ref{Fl2d}.

\begin{figure*}[t] 
    \centerline{
        \includegraphics[width=0.3\linewidth,clip]{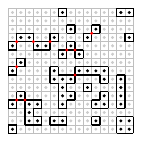}\hfill
        \includegraphics[width=0.37\linewidth,clip]{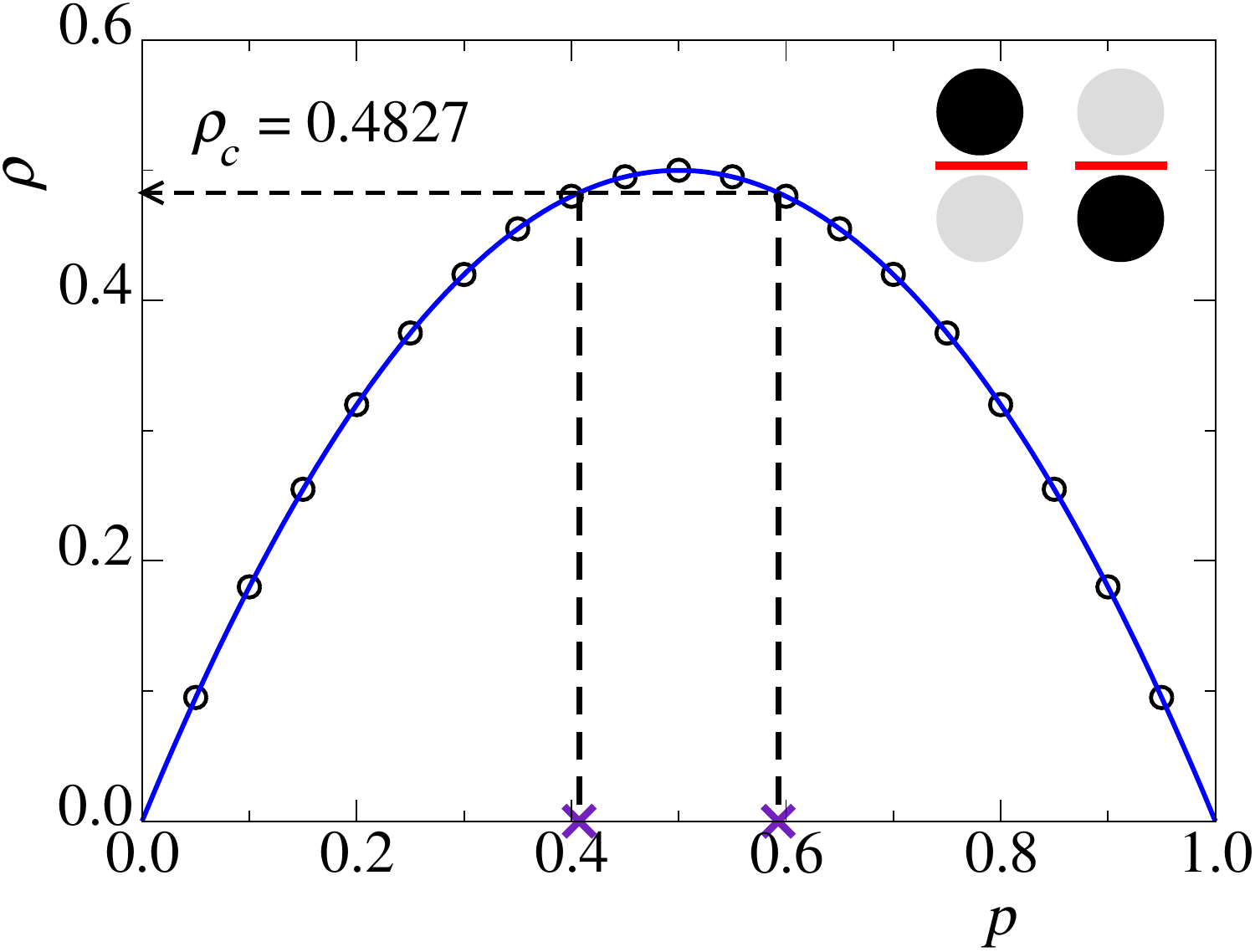}\hfill
        \includegraphics[width=0.3\linewidth,clip]{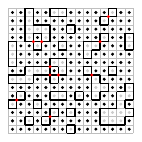}}
    \centerline{\hfill (a) $p=0.3$\hfill\hfill (b) \hfill\hfill (c) $p=0.7$\hfill}
    \caption{\label{morphology} (a) A typical morphology of occupied sites on a primal square lattice of size $16\times16$ and the perimeter bond clusters on its self-dual lattice, respectively,  for the site occupation probability $p=0.3$. Another morphology for $p=0.7$ is shown in $(c)$, where the perimeters appear around the unoccupied sites. In these morphologies, the grey bullets are unoccupied sites, and the black bullets are the occupied sites of the primal lattice. The thin grey lines between the lattice sites of the primal lattice represent the dual lattice, and the thick black lines represent the occupied bonds between two adjacent opposite sites. The red bullets on the dual lattice are the knots connecting two perimeter clusters. We presume an extra layer of unoccupied sites in (a) and an extra layer of occupied sites in $(c)$ on the boundary of the primal lattice. (b) Plot of bond density $\rho=2p(1-p)$ against $p$ is shown by a blue line. The black circles represent the measured bond density against $p$. Two configurations (interchanging positions of occupied and unoccupied sites) to generate a bond are shown in the right corner.}
\end{figure*}

All the links on the dual lattice are then verified, and bonds are placed on a dual lattice link only if they are found to be between two adjacent sites of different types (occupied-unoccupied) on the primal lattice. Thus, a bond on the dual lattice is separating a pair of occupied-unoccupied sites on the primal lattice.  Placing all such bonds, it is found that the occupied bonds on the dual lattice form perimeters of either the occupied site clusters or the unoccupied site clusters on the primal lattice. We call them perimeter bond clusters on the dual lattice. These bonds of mixed wet percolation represent the pores with zero capillary barriers, through which the two immiscible fluids are likely to flow in the two-component porous media in a capillary-dominated regime. We study the geometrical properties of these perimeter bond clusters. In Fig. \ref{morphology}(a) and (c), we present such perimeter bond clusters generated on the dual lattice of a $16\times16$ square primal lattice for a site concentration $p=0.3$ and $p=0.7$, respectively. 

It can be seen that the perimeter bond clusters of MWP are very different from random bond percolation (BP) clusters on a square lattice. The perimeter bond clusters of MWP are closed loops around the occupied (or unoccupied) sites without having a single open bond, in contrast to random bond percolation clusters. The smallest isolated cluster has four bonds, whereas in random bond percolation, a single-bond cluster is possible. Unlike random bond percolation, all clusters have an even number of bonds. The perimeters around two clusters of occupied (or unoccupied) sites on the primal lattice are connected by a knot shown by red bullets in Fig. \ref{morphology}(a) and (c). A lattice site on the dual lattice will be called a {\it knot} if all four occupied bonds meet the site. It can be noticed that the perimeter bond clusters are very similar to the boundary of up-spin and down-spin domains in a spin-$1/2$ Ising system. At $p=0.3$, the perimeters are around the clusters of occupied sites on the primal lattice. This is because the density of occupied sites is less than that of the unoccupied sites. Symmetrically, at $p=0.7$, the perimeters are mostly around the clusters of unoccupied sites on the primal lattice, as their concentration is less than that of the occupied sites. Since a bond in the dual lattice appears between a pair of adjacent occupied and unoccupied sites on the primal lattice, the density of bonds $\rho$ in terms of the site density $p$ is given by  
\begin{equation}
    \label{rhop}
    \displaystyle
    \rho=2p(1-p) \;,
\end{equation}
where $\rho=B/(2L^2)$, $B$ is the total number of occupied bonds in the square dual lattice of size $L\times L$ and the factor $2$ appears because a bond can also be generated by interchanging the positions of occupied-unoccupied sites, as shown in the inset of Fig. \ref{morphology}(b). In  Fig. \ref{morphology}(b), $\rho$ is plotted against $p$, the site occupation probability. The blue line represents Eq. \ref{rhop}, and the black circles represent the measured bond density against $p$. A good agreement is observed. As $p$ increases from zero, the value of $\rho$ increases from zero, and it reaches a maximum value $\rho_{\rm max}=1/2$ at $p=1/2$, then decreases and goes to zero at $p=1$. The bond density is symmetric about $p=1/2$. As $\rho$ is a quadratic function of $p$, for a given $\rho$, there will be two values for $p$. Therefore, if there is a percolation threshold at which a perimeter bond cluster spans the dual lattice for the first time at $p=p_c<1/2$, then there must be another threshold at $p=1-p_c>1/2$ beyond which no perimeter bond cluster will span the dual lattice. To determine the critical thresholds ($p_c$), we will check whether a perimeter bond cluster spanned or percolated the dual lattice for a given $p$. The mixed wet percolation (or perimeter bond percolation) in the dual lattice will occur within these two percolation thresholds. First, we will be studying the geometrical properties of the mixed wet percolation clusters on the dual lattice around these thresholds. The results will be compared with those of ordinary percolation. 

Later, we will describe random bond mixed wet percolation (RBMWP) for $p_c<p<(1-p_c)$ in which the bond in the dual lattice will be occupied with certain probability $\omega$ rather than with unit probability as described above and study the critical behaviour of the geometrical properties of random bond mixed wet percolation clusters.  The RBMWP model at $p_c$ and $1-p_c$ for $\omega=1$ corresponds exactly to the mixed wet percolation described above. 

MWP should not be confused with the so-called $AB$ percolation  \cite{wierman1989ab, wu2003ab} in which $A$ and $B$ sites are connected by a bond on the primal lattice, and no dual lattice is involved. Moreover, $AB$ percolation does not exhibit a percolation transition on the square lattice. It is rather known as anti-percolation on the square lattice.

\section{Scaling theory of geometrical quantities}
\label{scaling}
Since we will be studying the geometrical properties of mixed wet percolation or perimeter bond percolation clusters on the dual lattice, they should be studied in terms of the bond density $\rho$. However, there exists a functional relation, $\rho=2p(1-p)$,  between $\rho$ and  $p$. Hence, any mixed wet percolation cluster-related quantity in terms of the bond density $\rho$ can be obtained in terms of site concentration $p$. As $\Delta\rho=\rho-\rho_c\to 0$, a perimeter bond cluster property $\phi(\rho)$ either becomes singular or diverges. The scaling of $\phi(\rho)$ with $\Delta\rho$ can be written as
\begin{equation}
    \label{sc1}
    \displaystyle
    \phi(\rho)\sim \left|\Delta\rho\right|^{\pm q} \;,
\end{equation}
where $q$ is an exponent, and $\pm$ corresponds to a singularity ($+$) or to a divergence ($-$). Now, we obtain $\Delta\rho$ in terms of $\Delta p=p-p_c$ as
\begin{align}
    \label{sc2}
    \displaystyle
    \Delta\rho & = 2p(1-p)-2p_c(1-p_c)=a\Delta p-b\left(\Delta p\right)^2 \;,
\end{align}
where $a=2(1-2p_c)$ and $b=2$. The above equation can also be obtained by a Taylor series expansion of $f(p)=2p(1-p)$ around $p=p_c$. Now, the bond cluster property $\phi(\rho)$ can be obtained in terms of $\Delta p$ as
\begin{equation}
    \label{sc3}
    \displaystyle
    \phi(p)\sim \left|a\Delta p-b\left(\Delta p\right)^2\right|^{\pm q}=\left|\Delta p\right|^{\pm q}\left|\left(a-b\Delta p\right)\right|^{\pm q} \;,
\end{equation}
and the critical exponent associated with the leading singularity in $\phi(p)$ is found as
\begin{equation}
    \label{sc4}
    \displaystyle
    \lim\limits_{\Delta p\to 0}\frac{\ln\left|\phi(p)\right|}{\ln\left|\Delta p\right|} = \pm q \;.
\end{equation}
Hence, as $\Delta p\to 0$ (as well $\Delta\rho\to 0$), the scaling of $\phi(p)$ with $\Delta p$ is given by
\begin{equation}
    \label{sc5}
    \phi(p)\sim \left|\Delta p\right|^{\pm q} \;,
\end{equation}
where $q$ is the same scaling exponent with which $\phi(\rho)$ becomes singular or divergent as $\Delta\rho\to 0$. 

As the site percolation scaling theory is well described \cite{stauffer2018introduction, complexity}, we will provide the scaling relations for geometrical quantities related to mixed wet percolation clusters as a function of the control parameter $p$ around $p_c$ instead of the bond density $\rho$. A perimeter bond or mixed wet percolation cluster is characterized by the number of bonds $b$ present in that cluster and the cluster's radius of gyration $R_b$. The geometrical quantities of mixed wet percolation clusters at a site concentration $p$ on the primal lattice of size $L\times L$ can be defined in terms of the cluster number density $n_b(p)$ \cite{complexity,stauffer2018introduction,herrmann1990modelization}. The cluster number density $n_b(p)$ is defined as
\begin{equation}
\label{eq-4-1} 		
    \displaystyle
    n_b(p) =\frac{N_b(p)}{2L^2} \;,
\end{equation} 
where $N_b(p)$ is the number of perimeter bond clusters with $b$ bonds on the dual lattice. So, the perimeter bond cluster size distribution function can be written as
\begin{equation}
    \label{eq-4-2}
    \displaystyle
    n_b(p)=b^{-\tau}\mathcal{N}[(p-p_c)b^{\sigma}] \;,
\end{equation}
where $b$ is the number of bonds in a perimeter bond cluster and $\tau$ is an exponent. Hence, the order parameter $B_{\infty}(p)$, the probability of finding a bond in a spanning perimeter bond cluster is given by
\begin{equation}
    \label{eq-4-3}
    \displaystyle
    B_{\infty}(p) = \rho(p) - \sum_b{\vphantom{\sum}}'bn_b(p) \sim (p - p_c)^\beta \;,
\end{equation}
where the prime sum indicates exclusion of the largest cluster $b_{\rm{large}}$ and $\beta=(\tau-2)/\sigma$ is a critical exponent. The fluctuation in order parameter $\chi_b$ can also be obtained as,
\begin{equation}
    \label{eq-4-4}
    \displaystyle   
    \chi_b(p) = L^{d}\left[\langle B_{\infty}^2\rangle-\langle B_{\infty}\rangle^2\right]\sim |p - p_c|^{-\gamma} \;,
\end{equation}
where $\gamma=d\nu-2\beta=(3-\tau)/\sigma$ is another critical exponent. The correlation length $\xi_b$ is defined as 
\begin{equation}
    \label{eq-4-5}
    \displaystyle
    \xi_b^2(p) = 2\sum_b{\vphantom{\sum}}'R_b^2b^2n_b \bigg/ \sum_b{\vphantom{\sum}}'b^2n_b \sim |p-p_c|^{-2\nu} \;,
\end{equation}
where $\nu=(\tau-1)/d\sigma$ is the correlation length exponent, $d$ is the space dimension, and $R_b$ is the radius of gyration of a perimeter bond cluster of size $b$.

All the scaling relations above are valid for systems of large sizes ($L\to\infty$). However, most of the system sizes chosen in a simulation are finite, and the geometrical quantities $Q$ become a function of both $p$ and $L$. We must apply finite-size scaling (FSS) ansatz to study the critical behaviour of these geometrical quantities. If a geometrical quantity $Q$ scales as $Q\sim |p - p_c|^{-q}$ for a system of size $L\gg\xi$, then the FSS form of $Q(p,L)$ for the system size $L\ll\xi$ is given by,
\begin{equation}
    \displaystyle
    Q(p,L) = L^{q/\nu}\widetilde{Q}\left[(p - p_c)L^{1/\nu}\right] \;,
    \label{eq-4-6}
\end{equation}
where $q$ is a critical exponent, $\nu$ is the correlation length exponent and $\widetilde{Q}$ is the scaling function. Now, we provide FSS forms of some of the geometrical quantities we will use. The wrapping probability is the probability that a sample spans, and it is given by $W_b(p,L)=N_{\rm sp}/N$ where $N_{\rm sp}$ is the number of spanning samples out of $N$ samples. The FSS form of the wrapping probability $W_b(p,L)$ is given by
\begin{equation}
    \displaystyle
    W_b(p,L) = \widetilde{W}_b\left[(p-p_c)L^{1/\nu}\right] \;,
    \label{eq-4-7}
\end{equation}
as $W_b(p)$ is an analytic function, and no critical exponent is associated with it. $\widetilde{W}_b$ is the scaling function. The FSS form of the order parameter $B_\infty$ is 
\begin{equation}
    \displaystyle
    B_{\infty}(p,L) = L^{-\beta/\nu}\widetilde{B}_{\infty}\left[(p-p_c)L^{1/\nu}\right] \;,
    \label{eq-4-8}
\end{equation}
where $\widetilde{B}_{\infty}$ is the scaling function. The FSS of fluctuation in the order parameter $\chi_b$  is given by
\begin{equation}
    \displaystyle
    \chi_b(p,L)=L^{\gamma/\nu}\widetilde{\chi}_b\left[(p-p_c)L^{1/\nu}\right] \;,
    \label{eq-4-9}
\end{equation}
where $\gamma/\nu=d-2\beta/\nu$ and $\widetilde{\chi}_b$ is the scaling function. 

Note that the exponents $\tau,\sigma,\nu,\beta,\gamma$ described above are the critical exponents with which the perimeter bond cluster properties $\phi(\rho)$ become singular or divergent as $\Delta\rho\to 0$ in the dual lattice. 

\begin{figure}[ht]
    \centerline{\hfill
        \includegraphics[width=0.4\linewidth,clip]{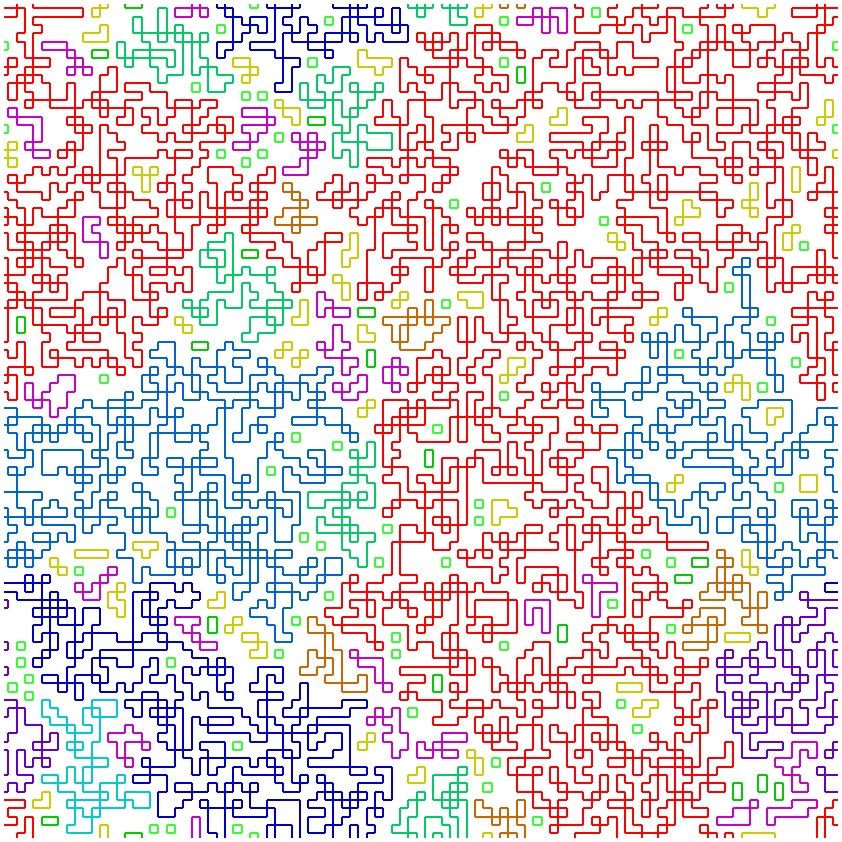}\hfill 
        \includegraphics[width=0.4\linewidth,clip]{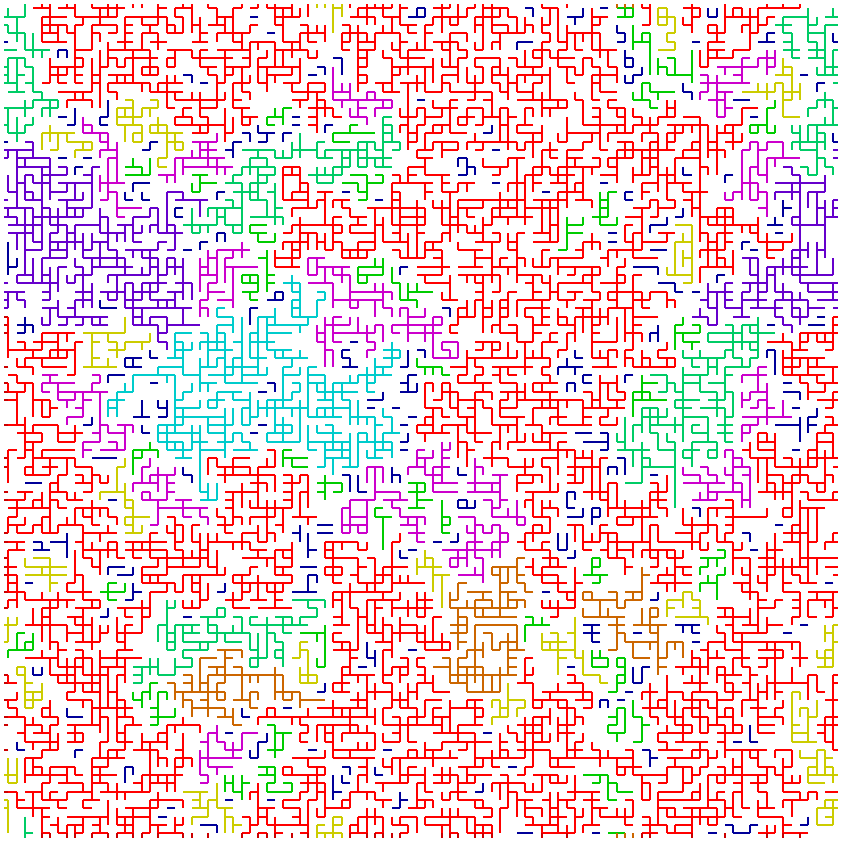}\hfill}
    \centerline{\hfill(a) $p=0.407$\hfill\hfill(b) $p=0.5$ \hfill}
    \caption{\label{clusters} (a) Morphology of the perimeter bond clusters generated on a $100\times 100$ dual lattice for site occupation probability $p=0.407$ on the primal lattice. (b) Morphology of bond clusters on a square lattice of the same size for bond occupation probability $p=0.5$. Different colours indicate different cluster sizes. The largest perimeter bond clusters are shown in red. Two clusters are found to be very different in nature.}
\end{figure}

The scaling behaviour of the geometrical quantities associated with the bond clusters of the RBMWP model should be studied in terms of $\Delta\omega=|\omega-\omega_c|$ where $\omega$ is the bond occupation probability. The scaling relations of the geometrical quantities in RBMWP in terms of $\Delta\omega$ will be the same as described above.

\section{Results and Discussion}
\label{rd}
In the following, we will first describe the simulation details of mixed wet percolation. We will then present the results and demonstrate different geometrical aspects of the model.

\subsection{Simulation}
\label{siml}
An extensive computer simulation of the above model has been performed on a square lattice and its dual lattice of different sizes $L\times L$, varying from $L=200$ to $2000$ in steps of $200$. Averaging of the geometrical quantities is made over $N=10^5$ samples. Periodic boundary conditions are applied in both directions. The burning algorithm \cite{hans} is used both in the horizontal and vertical directions to identify the spanning of the clusters. The Hoshen-Kopelman algorithm \cite{hk} is used to find the number of clusters $n_b$ with their sizes $b$ to determine the cluster size distributions and other geometrical properties of the cluster. In Fig.\ref{clusters} (a), we present a typical mixed wet percolation cluster configuration generated on a square dual lattice of size $100\times 100$ for site occupation probability $p=0.407$ on the primal lattice.  In Fig.\ref{clusters}(b), a typical random bond cluster configuration is shown for bond occupation probability $p=0.5$ on the square (primal) lattice of size $100\times 100$ for comparison to the MWP cluster configuration in Fig.\ref{clusters}(a). A particular colour is assigned to clusters of a certain range of cluster sizes. The ranges are fixed by logarithmic binning of sizes. The largest cluster is in red. It can be noticed that in Fig. \ref{clusters}(a) all clusters appear in the form of closed loops connected at certain corner points where four bonds meet. Moreover, in mixed wet percolation clusters (Fig.\ref{clusters}(a)), the red bonds (singly connected bonds) \cite{de1986multiscaling} are absent, whereas there are many such bonds in an ordinary random bond percolation cluster (Fig.\ref{clusters}(b)). The bob-link model \cite{stanley1977cluster} of percolation is then not applicable to mixed wet percolation, contrary to ordinary bond percolation. Thus, the perimeter bond clusters of mixed wet percolation are very different from the ordinary bond percolation clusters \cite{fan2021statistical}, where one finds open dangling ends and red bonds. It is, therefore, important to characterize the critical properties of the mixed wet percolation clusters. 

\begin{figure}[h]
    \centerline{\hfill
        \includegraphics[width=0.4\linewidth,clip]{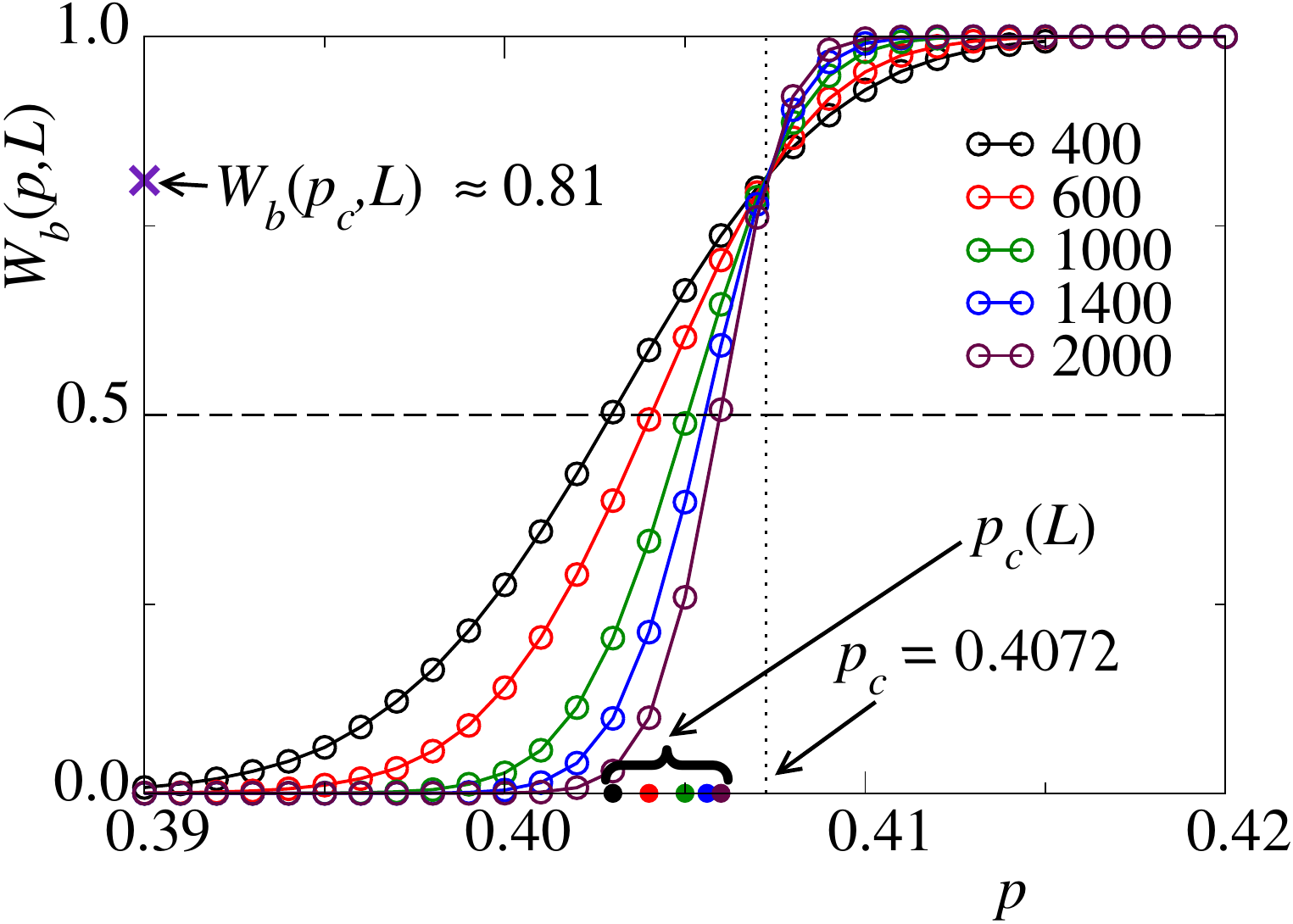}\hfill\hfill \includegraphics[width=0.4\linewidth,clip]{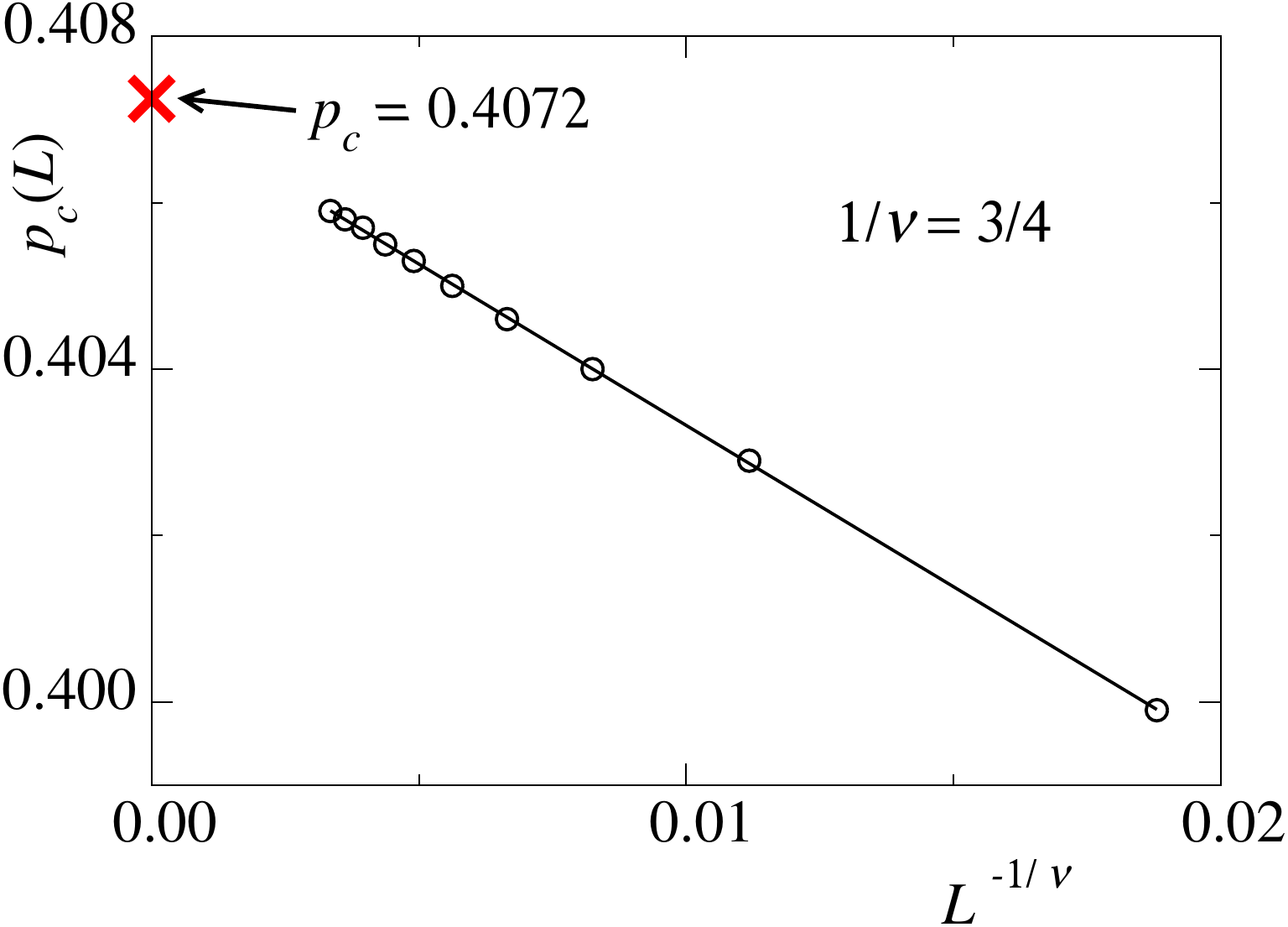}\hfill}
       \centerline{\hfill  (a) \hfill \hfill (b) \hfill }     
    \centerline{\hfill\includegraphics[width=0.4\linewidth,clip]{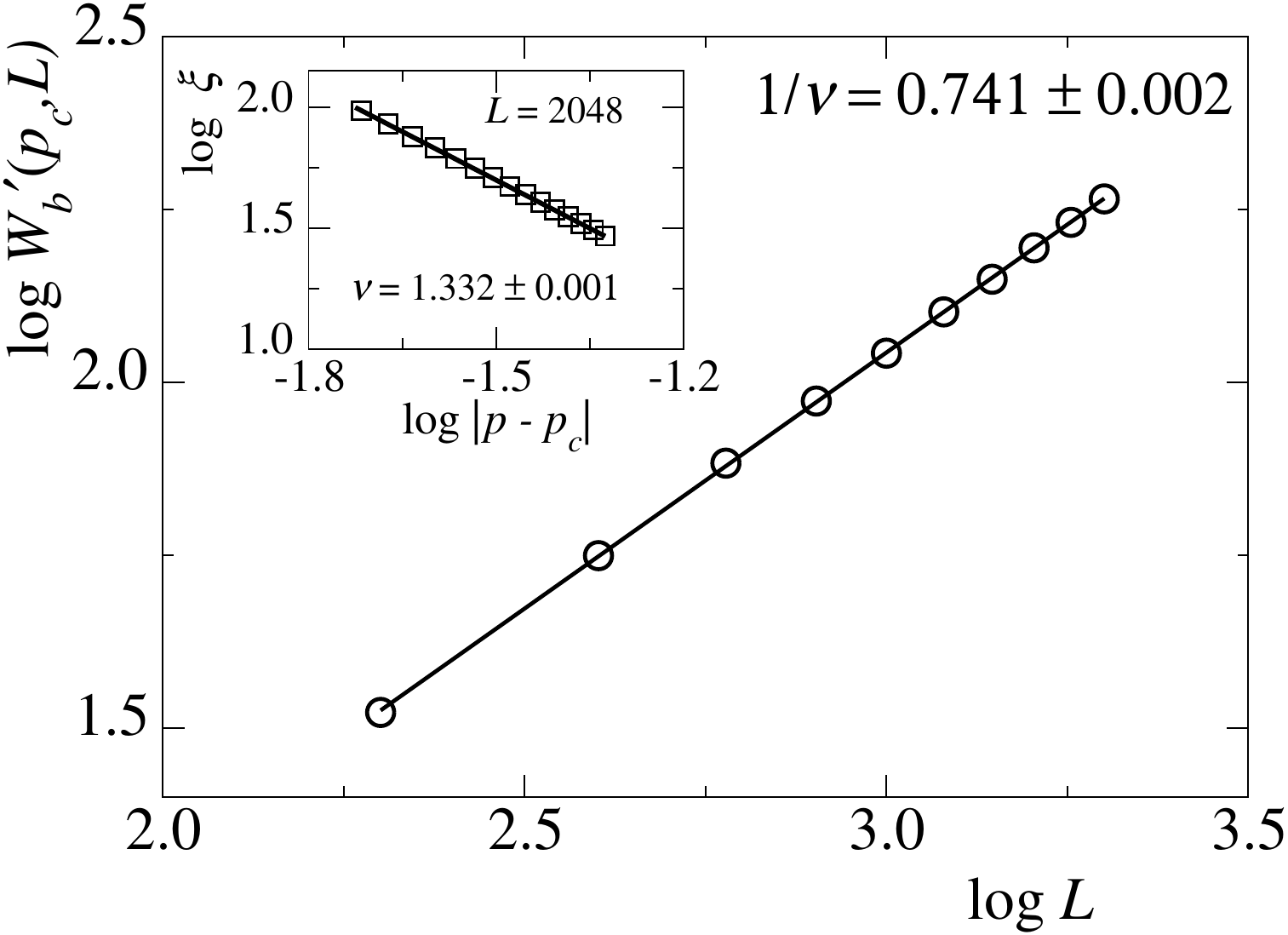}\hfill\hfill \includegraphics[width=0.4\linewidth,clip]{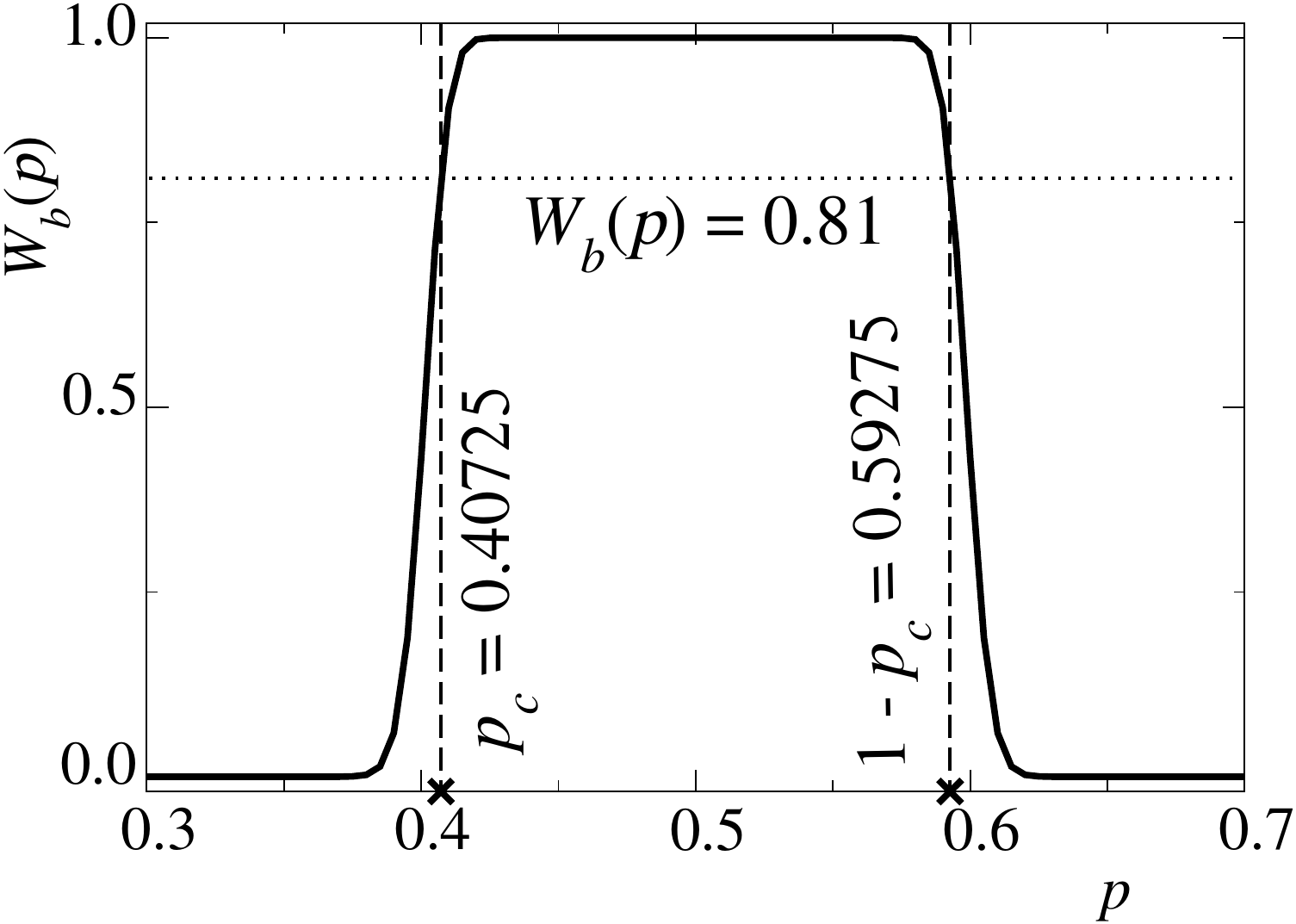}\hfill}
   \centerline{\hfill  (c) \hfill \hfill (d) \hfill }  
    \caption{\label{WP} (a) Plot of $W_b(p,L)$ versus $p$ for different $L$. All curves intercept at $p=p_c=0.4072$ as shown by a vertical dotted line. The dashed horizontal line represents $W_b(p, L)=1/2$. The intersection of $W_b(p,L)$ with $W_b(p,L) = 1/2$ occurs at the local threshold $p_c(L)$ of individual $L$. The coloured bullets on the $x$-axis represent $p_c(L)$ of the respective $L$. (b) Plot of $p_c(L)$ versus $L^{-1/\nu}$ taking $\nu=4/3$, the correlation length exponent of ordinary site percolation. The straight line represents the regression line. The intercepts with the $y$-axis, denoted by a red cross, yield $p_c=0.4072\pm 0.0034$. (c) Plot of $W'_b(p_c, L)$ versus $L$ in a double logarithmic scale. The data points are represented by circles. The straight line through the circles represents the regression line with slope $0.741\pm 0.002$. In the inset, the logarithm of correlation length $\log\xi$ is plotted against $\log|p-p_c|$ for $L=2048$ in the range $p<p_c$, shown by squares. The straight line through the squares is the regression line with slope $\nu=1.332\pm 0.001$. (d) Plot of $W_b(p, L)$ versus $p$ for the perimeter bond clusters on the dual lattice for a $1000\times 1000$ lattice. The black crosses on the $x$-axis indicate the thresholds $p_c$ and $1-p_c$.}
\end{figure} 

\subsection{Wrapping probability: $p_c$ and $\nu$}
\label{wp}
The critical threshold $p_c$ for $p<1/2$ is obtained by studying the warping probability $W_b(p,L)$ (Eq. \ref{eq-4-7}) varying $L$. In Fig. \ref{WP} (a), $W_b(p,L)$ of different $L$ are plotted against $p$. For a given $L$, data are taken varying $p$ from $0.39$ to $0.42$ in an interval of $0.0001$. For clarity, only thirty points are shown in a single graph out of an estimated three hundred points. Though we have estimated $W_b(p,L)$ for ten different values of $L$, we have kept plots for only five different values of $L$ in this figure for clarity. At $p=p_c$ the $W_b(p,L)=\widetilde{W}_b\left[0\right]$ is a constant. Thus, all the curves of $W_b(p,L)$ for different $L$ should intercept at $p=p_c$ as it is indicated by a vertical dotted line in  Fig. \ref{WP} (a). It should be noted that the value of $W_b(p_c,L)\approx 0.81$ at the threshold. This method provides a rough estimate of $p_c$, a more precise estimate of $p_c$ can be obtained by extrapolating the values of the local critical thresholds $p_c(L)$ of the systems of finite sizes $L$ to $L\to\infty$. For a finite system of size $L$, it is assumed that $W_b(p_c(L),L)=1/2$ at the threshold $p_c(L)$. Following the scaling form given in Eq. \ref{eq-4-7}, one can write $W_b(p_c(L),L)$ as
\begin{equation}
    \displaystyle
    W_b(p_c(L),L) = \widetilde{W}_b\left[(p_c(L)-p_c)L^{1/\nu}\right] =1/2 \;,
    \label{eq-3b-1}
\end{equation}
where $\nu$ is the correlation length exponent. Hence, one can obtain the local thresholds by taking the inverse of the scaling function $\widetilde{W}$ as 
\begin{equation}
    \displaystyle
    p_c(L)=p_c + CL^{-1/\nu} \;,
    \label{eq-3b-2}
\end{equation}
where $C=\widetilde{W}_b^{-1}\left[1/2\right]$ is a constant. To determine $p_c(L)$, a horizontal line $y=1/2$ (shown by a dashed line) is drawn to identify the intersection points with $W_b(p,L)$ for different $L$. Straight lines are fitted to $W_b(p,L)$ around $W_b(p,L)=1/2$. The point of intersection $(p_c(L),1/2)$ is then obtained by solving $y=1/2$ and the fitted line $y=y_0+\alpha{p}$ for each $L$, where $y_0$ and $\alpha$ are obtained through regression. The $p_c(L)$ values are marked by bullets (of the same colour) on the $x$-axis. 

In Fig. \ref{WP} (b), the values of $p_c(L)$ are plotted against $L^{-1/\nu}$ taking $\nu=4/3$, the correlation length exponent of the ordinary percolation. A straight line is then fitted through the data points. Since the points follow a straight line, it indicates the value of $\nu$ is taken rightly. From the intercept of the fitted line with the $y$-axis, we obtain the percolation threshold of mixed wet percolation at $p_c=0.4072\pm 0.0033$ indicated by a cross on the $y$-axis. The error includes the least square fit error and the error due to uncertainty in $p_c(L)$. The $p_c$ obtained here is close to the next-nearest-neighbour (nnn) site percolation threshold ($\approx 0.4072$) on the square lattice \cite{xun2021site}. Here, by the term next-nearest-neighbour (nnn), we refer to the $8$ neighbours surrounding a site on a square lattice. These include the $4$ nearest-neighbour sites (left, right, top, and bottom) and the $4$ diagonal sites at a distance $\sqrt 2$ times the lattice constant. Different studies have used varying terminologies for these neighbour classes. For example, in \cite{mg05}, they were denoted as SQ-1,2, whereas in \cite{nz00}, they were labeled as NN+NNN. Since this article does not discuss any additional nearest-neighbour sets, we will consistently use the term next-nearest-neighbour (nnn) to refer to the aforementioned eight neighbouring sites.

Note that $1-p_c({\rm OSP})=0.40725379$ where $p_c({\rm OSP})=0.59274621$ is the ordinary site percolation (OSP) \cite{nz00}. Thus, at $p_c$, along with mixed wet percolation in the dual lattice, nnn site percolation and breakdown of the infinite network of nearest-neighbour unoccupied sites occur on the primal lattice. It can be seen in Fig. \ref{morphology}(a) and (c) that a knot in mixed wet percolation in the dual lattice corresponds to an nnn bond on the primal lattice. The corresponding critical bond density $\rho_c$ on the dual lattice is $\rho_c=2p_c(1-p_c)\approx 0.4828$, and it is below $0.5$, the random bond percolation threshold on the square lattice.

Since, at $p=p_c$, the slope $W'_b(p_c,L)$, $(')$ indicates a derivative with respect to $p$, is given by
\begin{align}
    \label{nu2}
    \displaystyle
    W'_b(p_c,L)=cL^{1/\nu} \;,
\end{align}
where $c=\widetilde{W}'_b\left[0\right]$ is a constant, one can determine $1/\nu$ by measuring $W'_b(p_c,L)$ for different $L$. The slopes are measured right at the critical threshold $p_c$ by fitting a third-degree polynomial through ten data points around $p_c$ for each $L$. The values of $W'_b(p_c,L)$ are plotted against $L$ in a double logarithmic scale in Fig. \ref{WP}(c). By linear regression through the data points, we found $1/\nu=0.741\pm 0.002$, which is very close to the ordinary site percolation value of $1/\nu=0.75$. The value of the exponent $\nu$ is further verified by estimating the correlation length $\xi$ (Eq. \ref{eq-4-5}) as a function of $|p-p_c|$ on a lattice of size $L=2048$. In the inset, $\xi$ is plotted against $|p-p_c|$ in a double logarithmic scale. Through regression, we obtain $\nu=1.332\pm 0.001$. Thus, it shows that the correlation length exponent is the same as that of the ordinary site percolation, i.e.; $\nu=4/3$. Since the percolation criteria for the nnn percolation and perimeter bond percolation are the same, the wrapping probability $W(p,L)$ as a function $p$ for both problems will be the same. Hence, it is not surprising that the critical threshold $p_c$ and $1/\nu$ are the same for both problems. 

Now, we investigate the general form of the wrapping probability in terms of $p$ on a large single lattice. In Fig. \ref{WP}(d), $W_b(p)$ for $L=1000$ is plotted against $p$, varying $p$ from $0.3$ to $0.7$. It can be seen that the $W_b(p)$ is symmetric around $p=1/2$. The horizontal dotted line represents the value of $W_b(p_c,L)$. This line intercepts the first dashed vertical line ($p_c=0.40725$) at the threshold point. It also intercepts $W_b(p)$ on the other side ($p>1/2$) at another threshold, which is indicated by another dashed line $1-p_c=0.59275$. Two black crosses on the $x$-axis indicate these threshold values. Note that the second threshold, $1-p_c=0.59275$, is the ordinary site percolation  threshold $p_c({\rm OSP})$ \cite{stauffer2018introduction}. Then the first $p_c$, is $p_c=1-p_c({\rm OSP})=0.40725$, the threshold of nnn site percolation.  

In the $L\to\infty$ limit, the wrapping probability $W_b(p, L)$ in terms of $p$ can be expressed as a product of two Heaviside step functions having steps at $p_c=1-p_c({\rm OSP})$ and $1-p_c=p_c({\rm OSP})$. It can be expressed as
\begin{align}
    \displaystyle   
    W_b(p,\infty) &=H(p-p_c)H\left[(1-p_c)-p\right] \;,
    \label{eq-5-1}
\end{align}
where $H(p-p_c)$ is zero for $p<p_c$ and one for $p>p_c$ whereas $H\left[(1-p_c)-p\right]$ is one for $p<(1-p_c)$ and zero for $p>(1-p_c)$. 

\begin{figure}[t]
    \centerline{\hfill\includegraphics[width=0.4\linewidth]{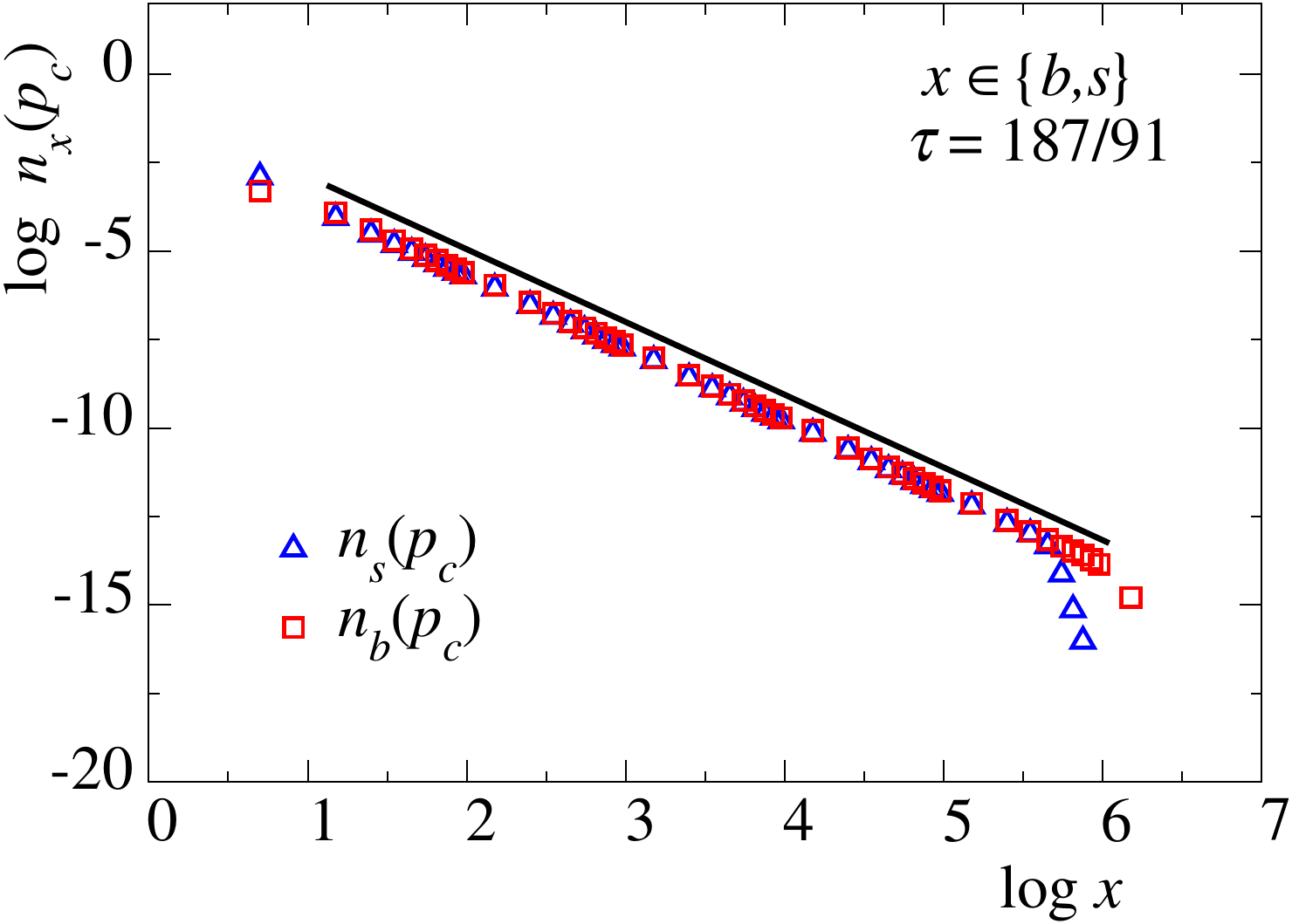}\hfill\hfill
        \includegraphics[width=0.4\linewidth]{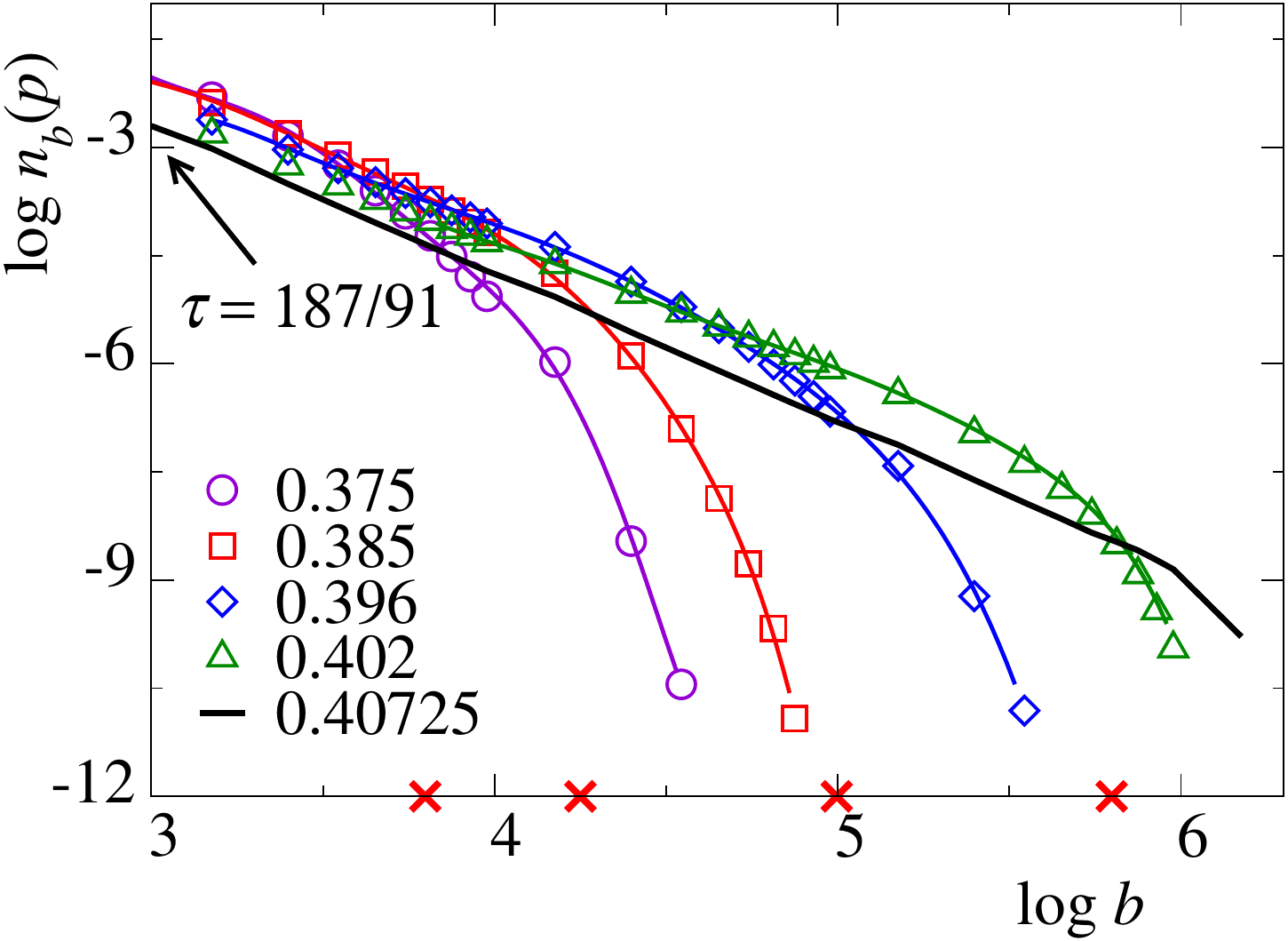}\hfill}
    \centerline{\hfill (a)\hfill\hfill (b)\hfill}    
   \centerline{\hfill\includegraphics[width=0.4\linewidth]{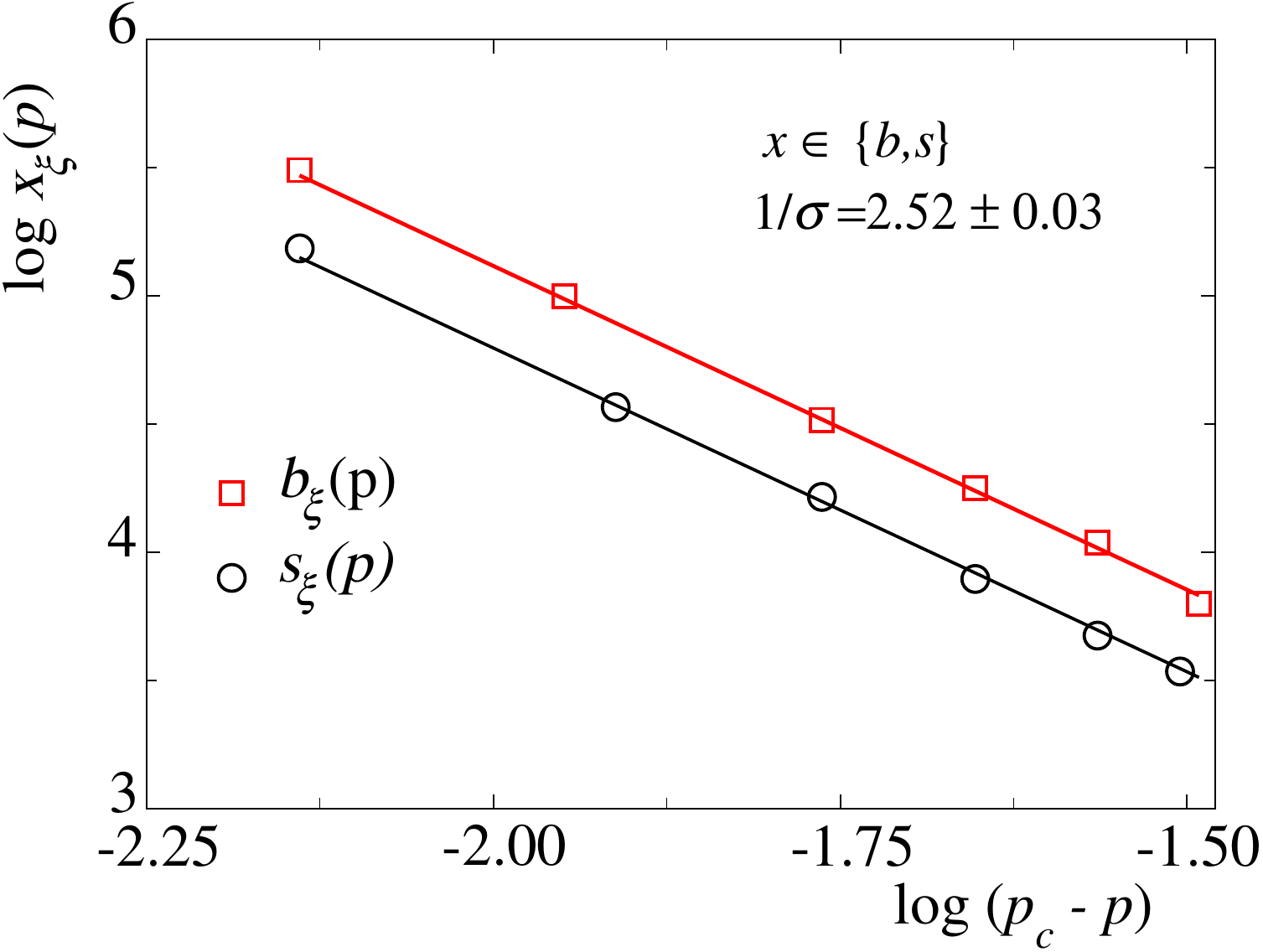}\hfill\hfill \includegraphics[width=0.4\linewidth]{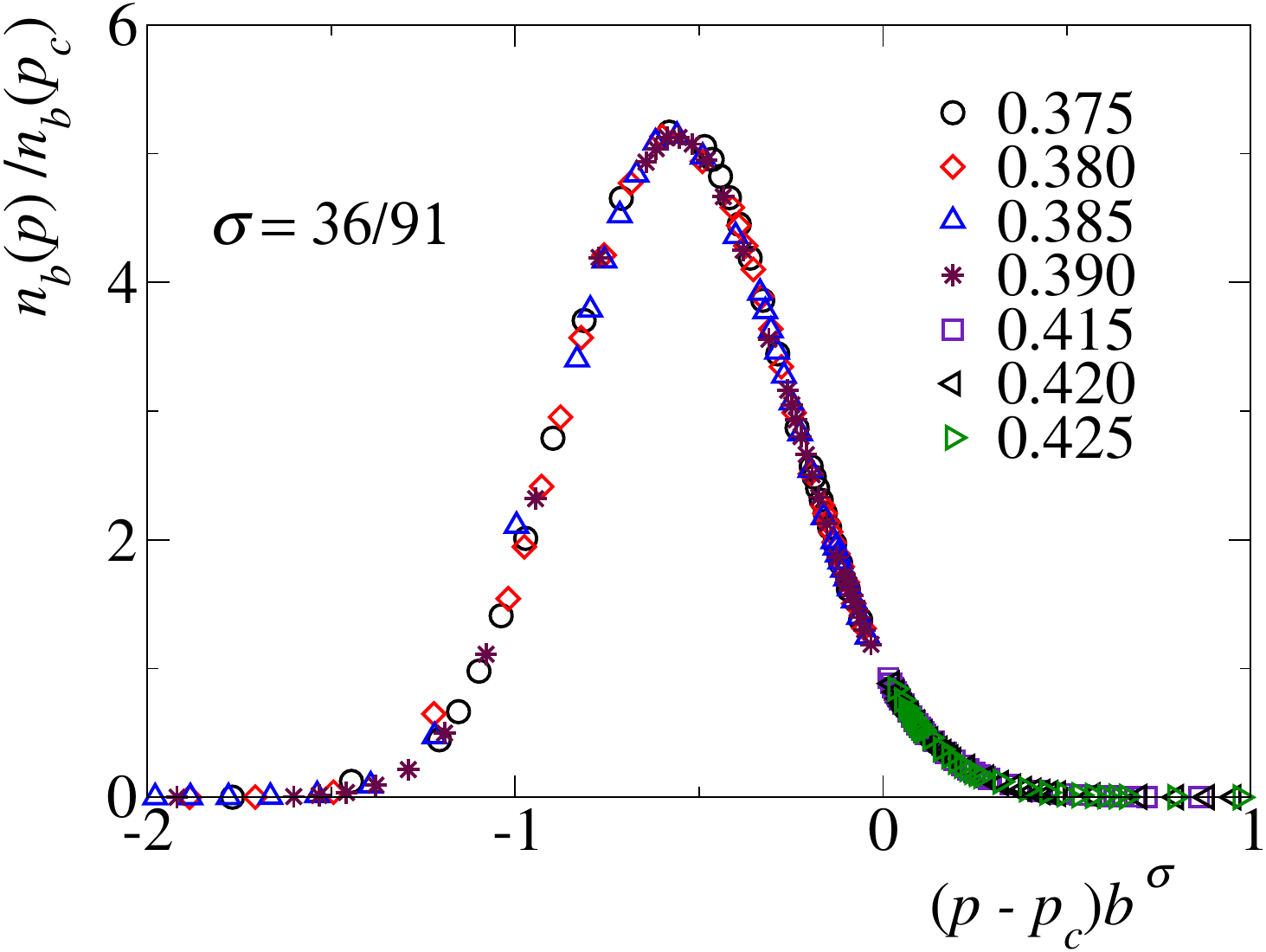}\hfill}
       \centerline{\hfill (c)\hfill\hfill (d)\hfill} 
    \caption{\label{cluster_size} (a) Plot of cluster size distribution of perimeter bond clusters $n_b(p_c)$ against the size of the perimeter bond cluster $b$ and cluster size distribution of nnn site clusters $n_s(p_c)$ against nnn cluster size $s$ in a double logarithmic scale. (b) Plot of cluster size distribution of perimeter bond clusters $n_b(p)$ against perimeter cluster size $b$ in a double logarithmic scale for various values of $p$, indicated by different symbols, for a system size of $L = 2048$. (c) Plot of $b_\xi$ and  $s_\xi$ against $p-p_c$ in a double logarithmic scale. The value of the critical exponent $\sigma = 2.52 \pm 0.03$ was determined. (d) Plot of the collapse of ratio of cluster size distributions $n_b(p)/n_b(p_c)$ against $(p-p_c)b^\sigma$ for various values of $p$ at lattice size $L = 2000$.}
\end{figure} 

\subsection{Cluster size distribution}
Though the correlation length exponent $\nu$ remains the same as that of random bond percolation, it is important to verify the cluster size distribution of the perimeter bond clusters. The cluster size is defined as the number of bonds $b$ present in a perimeter bond cluster. The perimeter bond cluster size distribution $n_b(p)=b^{-\tau}\mathcal{N}[(p-p_c)b^\sigma]$ (Eq. \ref{eq-4-2}) becomes a power law at $p=p_c$ as it is given by
\begin{equation}
    \label{eq-3-19}
    \displaystyle
    n_b(p_c)\sim b^{-\tau} \;,
\end{equation}
as $\mathcal{N}[0]$ is a constant. In Fig. \ref{cluster_size}(a), the cluster size distribution $n_b(p_c)$  of the perimeter bond cluster is plotted against the cluster size $b$ for the system of size $L=2000$ in the double logarithmic scale shown by the red squares. For comparison, the size distribution of the nnn site percolation clusters $n_s(p_c)\sim s^{-\tau}$ is shown by blue triangles in Fig. \ref{cluster_size}(a). The slopes of the straight part of both the distributions are found to be $2.054\pm 0.002$. It is very close to the exponent $\tau=187/91$ as that of ordinary site percolation, shown by a black line as a guide to the eye. Note that the distributions $n_b(p_c)$ and $n_s(p_c)$ are two independent distributions. However, they only differ in the first bin and in the last few bins, though the total number of site and perimeter clusters are the same. In the first bin ($s\text{ or } b\in 1-10$), though there are all possible nnn site percolation clusters, the perimeter bond clusters appear in even numbers starting from $b=4$ only. The largest bin of $n_b(p_c)$ is greater than that of $n_s(p_c)$. This indicates that the perimeter of a site cluster can be more than the size of the site cluster.  

Though the distribution is fairly power law for $p=p_c$, it deviates from the power law for $p<p_c$. The perimeter bond cluster size distributions for $p<p_c$ are plotted in Fig. \ref{cluster_size}(b) in double logarithmic scale. There is a strong deviation from the power law as $p$ is smaller than $p_c$. For a given $p$, if such a crossover occurs at a certain value of the perimeter bond cluster size $b_\xi(p)$, the cluster size distribution can be written in terms of $b_\xi$ as in \cite{sorensen}
\begin{equation}
    \label{eq-3-20}
    \displaystyle
    n_b(p)=n_b(p_c)F\left(\frac{b}{b_\xi}\right) \;,
\end{equation}
where $n_b(p_c)\sim b^{-\tau}$. Here $b_\xi$ scales as
\begin{equation}
    \label{eq-3-21}
    b_\xi=b_0\left|p-p_c\right|^{-1/\sigma} \;,
\end{equation}
where $\sigma$ is an exponent. We have measured $b_\xi$ from the intersection of the curve $\log(n_b(p))$ with the line $\log(n_b(p_c))$. To find the intersection point, we fit a higher order polynomial through $\log{(n_b(p))}$ around the point of intersection and find the solution with the line of $\log(n_b(p_c))$. Red crosses mark the $\log b_\xi$ values on the $x$-axis. In Fig. \ref{cluster_size}(c), $b_\xi$ is plotted against $|p-p_c|$ in double logarithmic scale and found that $b_\xi$ follows a power law with an exponent $1/\sigma=2.52\pm 0.03$ which is very close to the percolation value $1/\sigma=91/36\approx 2.53$. The corresponding quantity in nnn site percolation $s_\xi$ is also plotted for comparison. It seems the value of $\sigma$ is the same as that of the ordinary percolation.

Now, we verify the scaling function form of the cluster size distribution. The ratio $n_b(p)/n_b(p_c)$ can be written as
\begin{equation}
    \label{eq-3-22}
    \displaystyle
    \frac{n_b(p)}{n_b(p_c)}=\frac{\mathcal{N}[z]}{\mathcal{N}[0]},
\end{equation}
is a normalized scaling function of the scaled variable $z=(p-p_c)b^\sigma$. The data for different $p$ for a given $L$ should collapse onto a single curve if $n_b(p)/n_b(p_c)$ is plotted against $z$. In Fig. \ref{cluster_size}(d), the ratio $n_b(p)/n_b(p_c)$ is plotted against the scaled variable $z=(p-p_c)b^\sigma$ for several values of $p$ for the system of $L=2000$ taking $\tau=187/91$ and $\sigma=36/91$ as that of the ordinary site percolation . It can be seen that data for several $p$ have collapsed to a single curve with the $\tau$ and $\sigma$ of the ordinary site percolation. It can be noted that at $p=p_c$, one finds $z=0$ and that corresponds to $\mathcal{N}[0]/\mathcal{N}[0]=1$ on the $y$-axis. In terms of the parameter $\rho$, $n_b(\rho)$ is indifferent from $n_b(p)$. However, the normalized plot of $n_b(\rho)/n_b(\rho_c)$ against the scaled variable $z=(\rho-\rho_c)b^{\sigma}$ will have the same height but squeezed width.   

\begin{figure}[h]
    \centerline{\includegraphics[width=0.4\linewidth,clip]{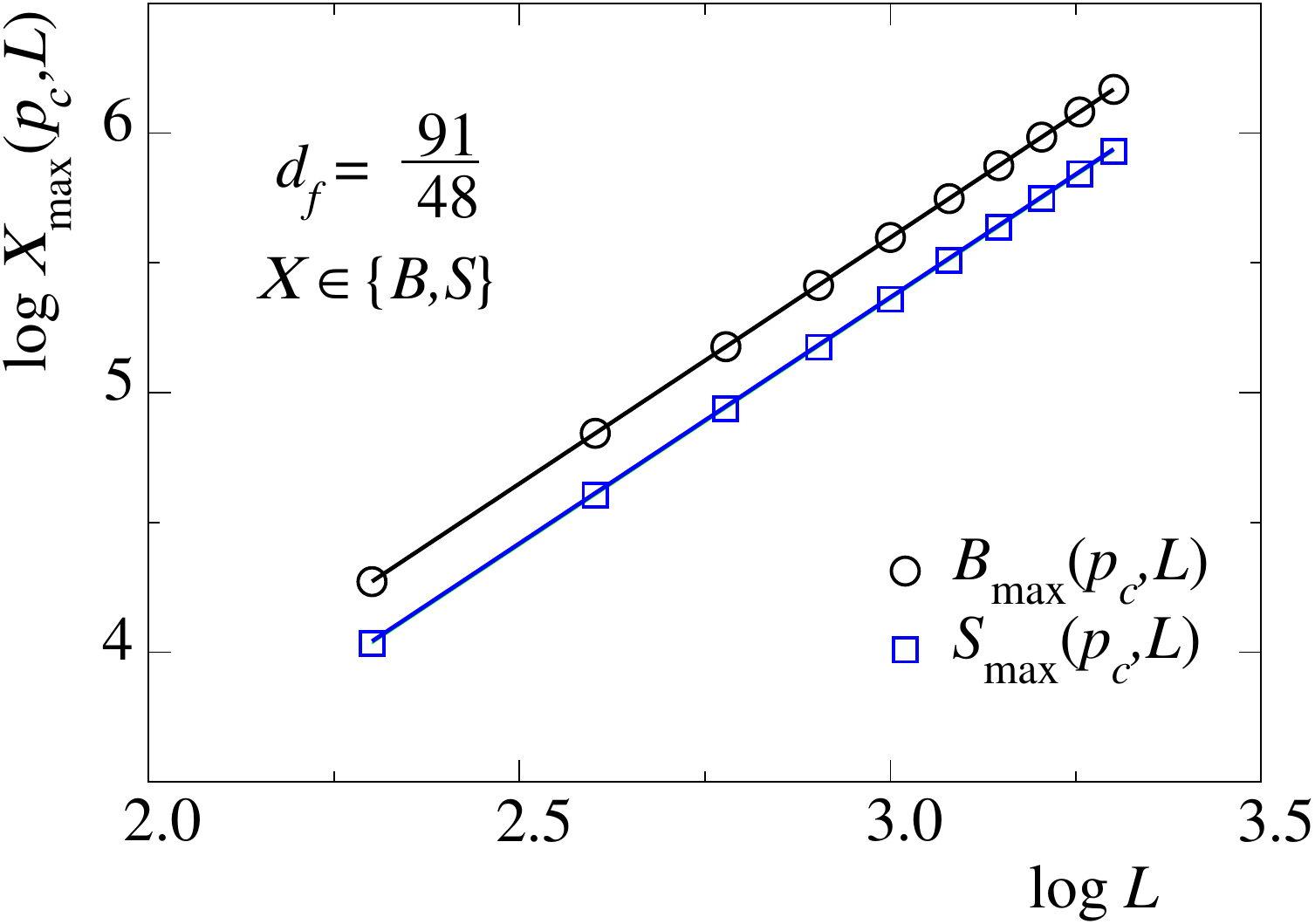}}
    \caption{\label{df} Plot of mass of the spanning perimeter bond cluster $B_{\rm max}$ (on the dual lattice) against $L$ in black circles and plot of mass of the nnn spanning site cluster $S_{\rm max}$ (on the primal lattice) against $L$ in blue squares. The black and the blue lines of slope $91/48$ are guides to the eye.}
\end{figure} 

\subsection{Fractal dimension}
Since the exponents $\tau$ and $\sigma$ of the cluster size distribution are the same as those of the ordinary site percolation, all other exponents related to different moments of the $n_b(p)$ would be the same. However, it would be interesting to see whether the fractal dimension will be that of the percolation hull \cite{ziff1986test} or that of the ordinary site percolation clusters. We present here an independent measurement of the fractal dimension of the spanning perimeter bond clusters. A typical spanning perimeter bond cluster (in red) is shown in Fig.\ref{clusters}(a). The spanning cluster looks very different from a spanning ordinary bond percolation cluster in Fig.\ref{clusters}(b). However, the cluster has holes of all possible sizes. It is tortuous and looks self-similar. The mass of the spanning perimeter bond clusters at $p=p_c$ is supposed to scale with the system size $L$ as $B_{\rm max}(p_c,L)\sim L^{d_f}$. Note that $B_{\rm max}(\rho_c,L)$ and $B_{\rm max}(p_c,L)$ are the same, as they correspond to the same critical point. In Fig. \ref{df}, the average mass $B_{\rm max}$ of the spanning cluster (shown in black circles) is plotted against $L$ on a log-log scale. The black straight line of slope $d_f=91/48$ is a guide to the eye. The data points follow the black straight line quite nicely. The blue squares represent the average mass $S_{\rm max}$ of the spanning clusters of occupied sites on the primal lattice. Those site clusters represent the nnn site percolation spanning clusters. The blue line is parallel to the red line and has a slope $d_f=91/48$. Thus, the fractal dimension $d_f$ of the mixed wet percolation spanning cluster and that of the nnn site percolation are the same as that of the ordinary (nn) site percolation spanning cluster. It can be seen that $B_{\rm max}(L)>S_{\rm max}(L)$ as $\rho_c>p_c$. Therefore, even if there is a mass difference at a given $L$, both of them scale with the same fractal dimension. Since $d_f$ and other exponents are the same as those of percolation, all the hyper-scaling relations are satisfied for the perimeter bond cluster percolation. It is surprising that the fractal dimension of the perimeter loops of MWP is different from the hull fractal dimension ($d_h = 7/4$) and identical to that of the bulk ($d_f = 91/48$). A closer look at the infinite cluster of MWP reveals that it is essentially made of several perimeter loops connected by knots and forms a giant cluster that covers the whole lattice instead of being the perimeter of a single large cluster of occupied (unoccupied) sites. The giant perimeter loop cluster has almost the same tortuosity or randomness as that of the percolation cluster, and that results in a bulk fractal dimension.

\begin{figure}[h]
    \centerline{\includegraphics[width=0.4\linewidth,clip]{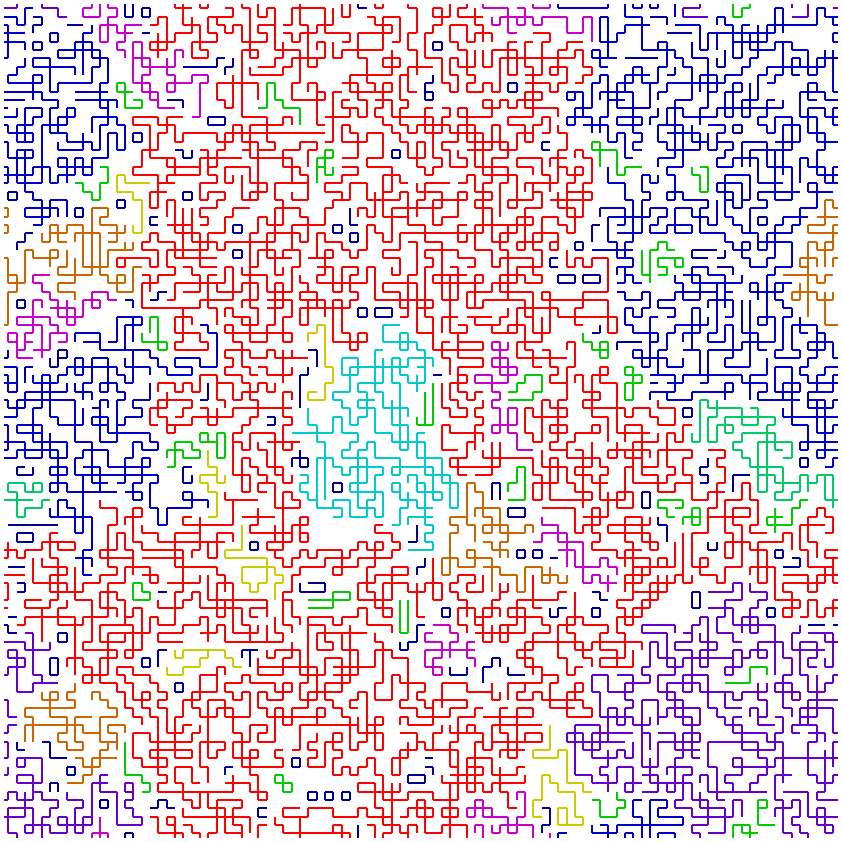}}
    \caption{\label{morphology_random} Typical morphology of random bond perimeter clusters on a square lattice, on a lattice of size $100\times100$ at the site occupation probabilities $p= 0.50$ and bond occupation probability $\omega =0.938$.}
\end{figure} 

\subsection{Random bond mixed wet percolation (RBMWP) model and results}
\label{rbmwp}
In the mixed wet percolation model, discussed above in Section \ref{model}, the randomness of the system is in the occupation of sites in the primal lattice. This randomness was borrowed by the perimeter bond clusters when the bonds were placed on the dual lattice between a pair of occupied and unoccupied sites on the primal lattice with unit probability. There was no inherent randomness of the perimeter bond cluster. Now, we would like to introduce randomness to the perimeter bond clusters. In RBMWP, the bonds on the dual lattice between a pair of occupied and unoccupied sites are occupied with a probability $(\omega)$ rather than a unit probability. In that case, all the clusters may not be the perimeters of the site clusters on the primal lattice. The bond clusters may have dangling ends, unlike in the mixed wet percolation model, where the network consists of closed loops only. At a given $p$, between $p_c$ and $1-p_c$, the system is studied by varying the bond occupation probability $\omega$. It is expected that at a critical threshold $(\omega_c)$, the system will undergo a percolation transition. In this section, we will explore the critical behaviour of the RBMWP model.

In Fig. \ref{morphology_random}, we present a typical morphology of RBMWP on $100\times100$ square dual lattice at $p = 0.5$ and $\omega = 0.935$. The RBMWP clusters do not look very different from the mixed wet percolation clusters shown in  Fig. \ref{clusters}(a) except for some dangling bonds. The clusters in RBMWP need not always be a combination of closed loops.
 
\begin{figure}[h] 
    \centerline{\hfill
        \includegraphics[width=0.33\linewidth,clip]{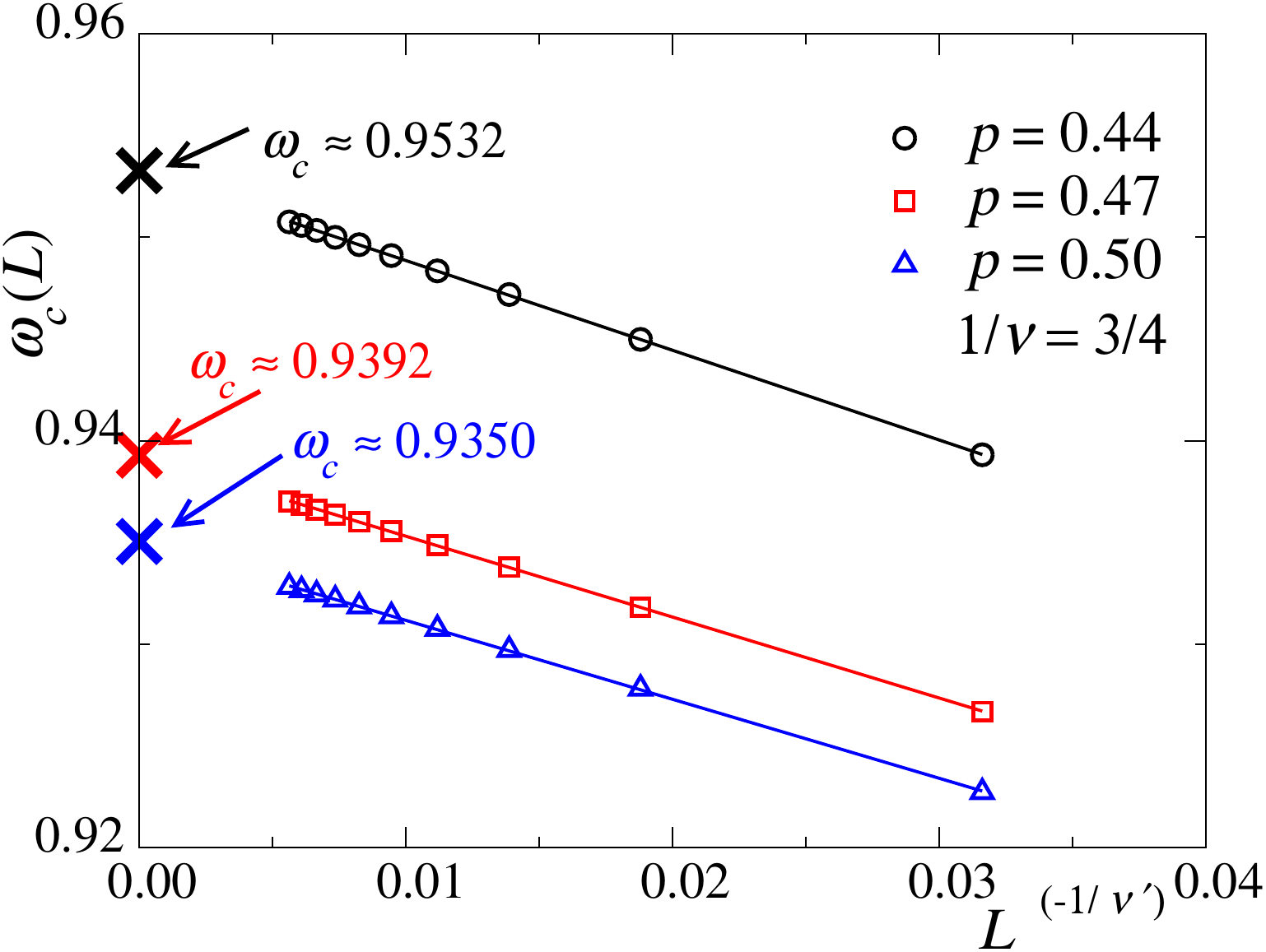}\hfill
        \includegraphics[width=0.33\linewidth,clip]{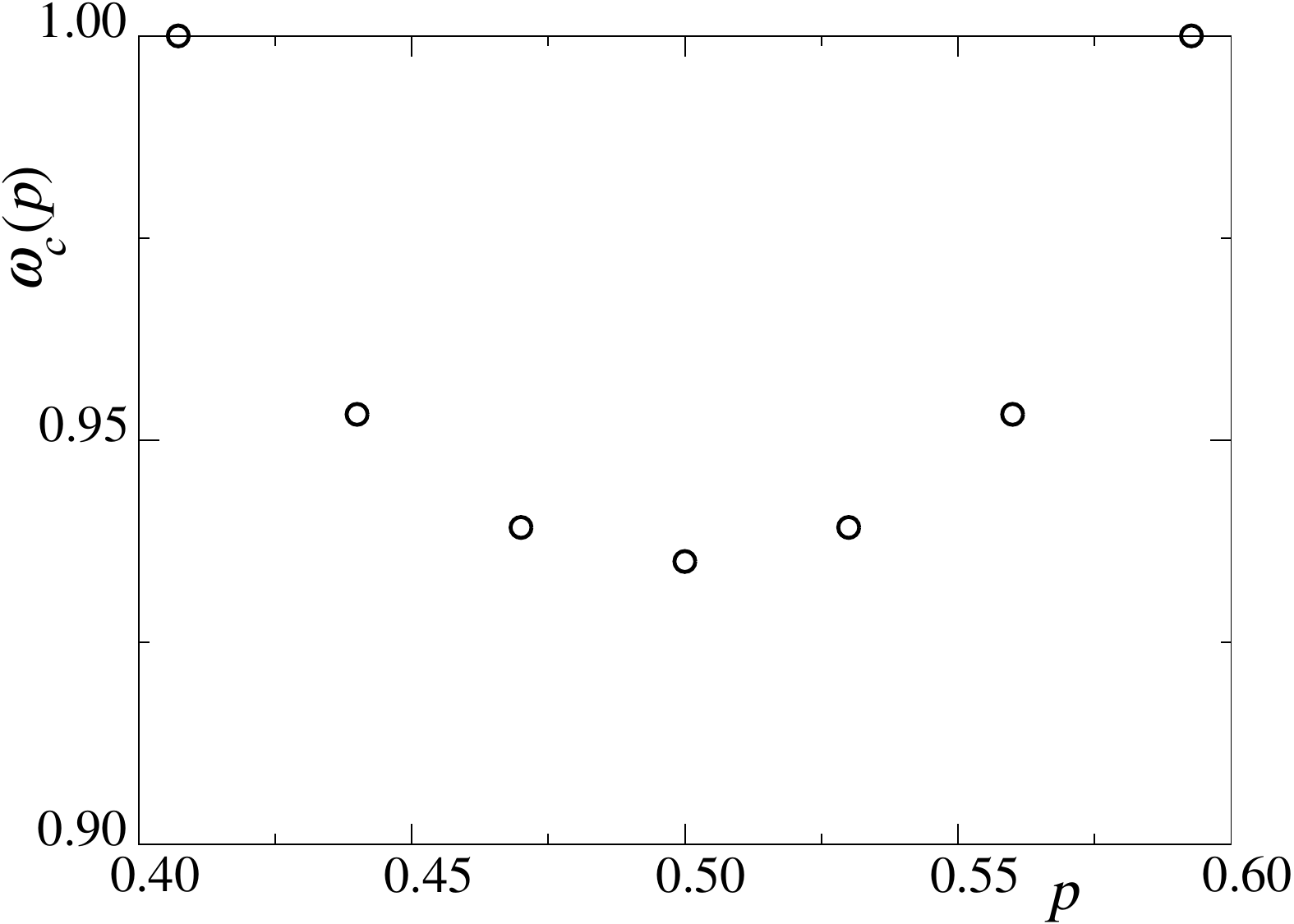}\hfill \includegraphics[width=0.31\linewidth,clip]{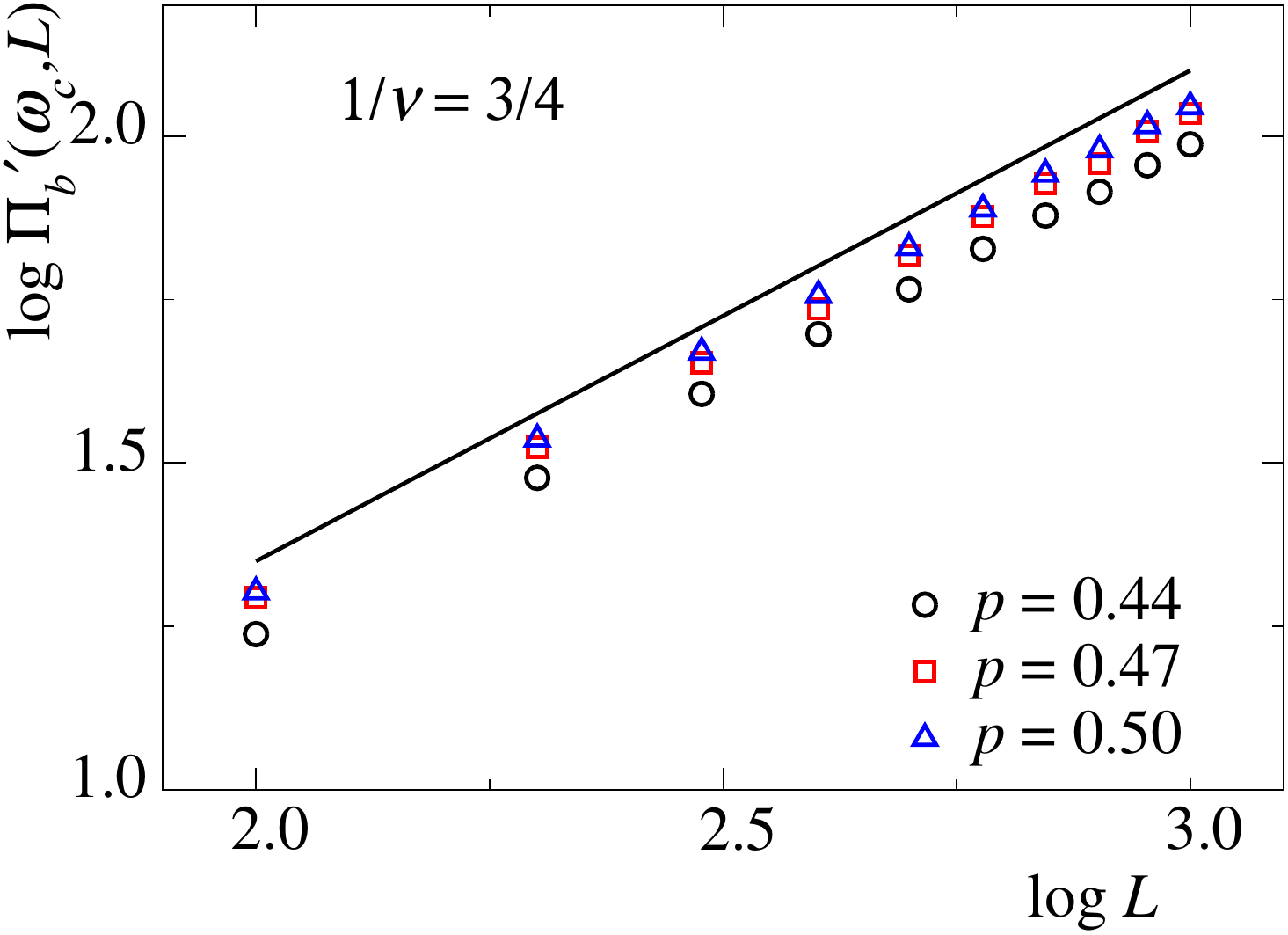}\hfill}
 \centerline{\hfill (a) \hfill \hfill (b) \hfill\hfill (c) \hfill }               
    \caption{\label{pc_random} (a) Plot of $\omega_c(L)$ versus $L^{-1/\nu'}$ taking $\nu'=4/3$, the correlation length exponent of percolation. This is done for values of $p = 0.44, 0.47$ and $0.50$. (b) Plot of $\omega_c$ for various values of $p$ between $p_c$ and $1 - p_c$. (c)  Plot of $\Pi'_b(\omega_c,L)$ versus $L$ in double logarithmic scale for $p = 0.44,0.47$ and $0.50$. The correlation length exponent is found as $1/\nu'\approx 0.74\pm 0.004$ for all three cases.}
\end{figure} 

The methods used to study RBMWP are the same as those used to study the mixed wet percolation, replacing $p$ by $\omega$. All geometrical quantities will be studied carrying $\omega$ for each $p$. The wrapping probability of bond clusters in RBMWP, $\Pi_b(\omega,L)$, is then given by
\begin{equation}
    \label{Pi1}
    \displaystyle
    \Pi_b(\omega,L) = \widetilde{\Pi}_b\left[(\omega-\omega_c)L^{1/\nu'}\right] \;,
\end{equation}
where $\widetilde{\Pi}_b$ is the scaling function, $\omega_c$ is the critical threshold at a given $p$ and $\nu'$ is the correlation length exponent for RBMWP. We first determine the critical threshold of the bond occupation probability $\omega_c$ at which the random bond clusters percolate. Since the systems are finite in size, we define the local thresholds as 
\begin{equation}
    \label{Pi2}
    \displaystyle
    \Pi_b(\omega_c(L),L) = \widetilde{\Pi}_b\left[(\omega_c(L)-\omega_c)L^{1/\nu'}\right] = 1/2 \;,
\end{equation}
where $\omega_c(L)$ is the critical threshold for a system of size $L$. Inverting the Eq. \ref{Pi2} we obtain $\omega_c(L)$ as,
\begin{equation}
    \displaystyle
    \omega_c(L) = \omega_c + AL^{-1/\nu'} \;,
    \label{5-b1}
\end{equation}
where $A=\widetilde{\Pi}_b^{-1}\left[1/2\right]$ is a constant. In the vicinity of $\Pi_b(\omega,L)=1/2$ a straight line is fitted and the intersection of the fitted line with the line $y =1/2$ yields the thresholds $\omega_c(L)$. In Fig. \ref{pc_random}(a) the values of $\omega_c(L)$ are plotted  against $L^{1/\nu'}$ taking the correlation length exponent $\nu'=4/3$ for $p=0.44,0.47$ and $0.50$. A straight line is then fitted through them, and from the intercept of the fitted line with the y-axis, the critical threshold $\omega_c$ was determined. The $\omega_c$ obtained for $p = 0.44, 0.47 \text{ and } 0.50$ are $0.9532 \pm 0.0031,0.9392 \pm 0.0022 \text{ and } 0.0.9350 \pm 0.0025 $ respectively. For every value of $p$ between $p_c$ and $1-p_c$, there exists a value of $\omega_c$ for which the system is critical. In Fig. \ref{pc_random}(b), the values of $\omega_c$ are shown for a few values of $p$ between the two thresholds. It can be noted that it is the lowest at $p=1/2$.

To determine the correlation length exponent $\nu$, we take a derivative of $\Pi_b(\omega,L)$  at $\omega=\omega_c$ and it is given by,
\begin{equation}
    \displaystyle
    \Pi'_b(\omega_c,L)=aL^{1/\nu'} \;,
    \label{5-b2}
\end{equation}
where $a=\widetilde{\Pi}'_b[0]$ and the prime ($'$) indicates the derivative with respect to $\omega$. The derivatives are measured at the critical thresholds $\omega_c$ by fitting a third-degree polynomial around $\omega_c$ for each $L$. In Fig. \ref{pc_random}(c), $\Pi'_b$ is plotted against $L$ in a double logarithmic scale. The slope that indicates $1/\nu'$ was determined by linear regression and was found to be $1/\nu'=0.740\pm 0.004$. That is very close to that of the ordinary percolation one, {\em i.e.}; $\nu'=\nu=4/3$.

\begin{figure}[ht]
    \centerline{\hfill
    \includegraphics[width=0.33\linewidth,clip]{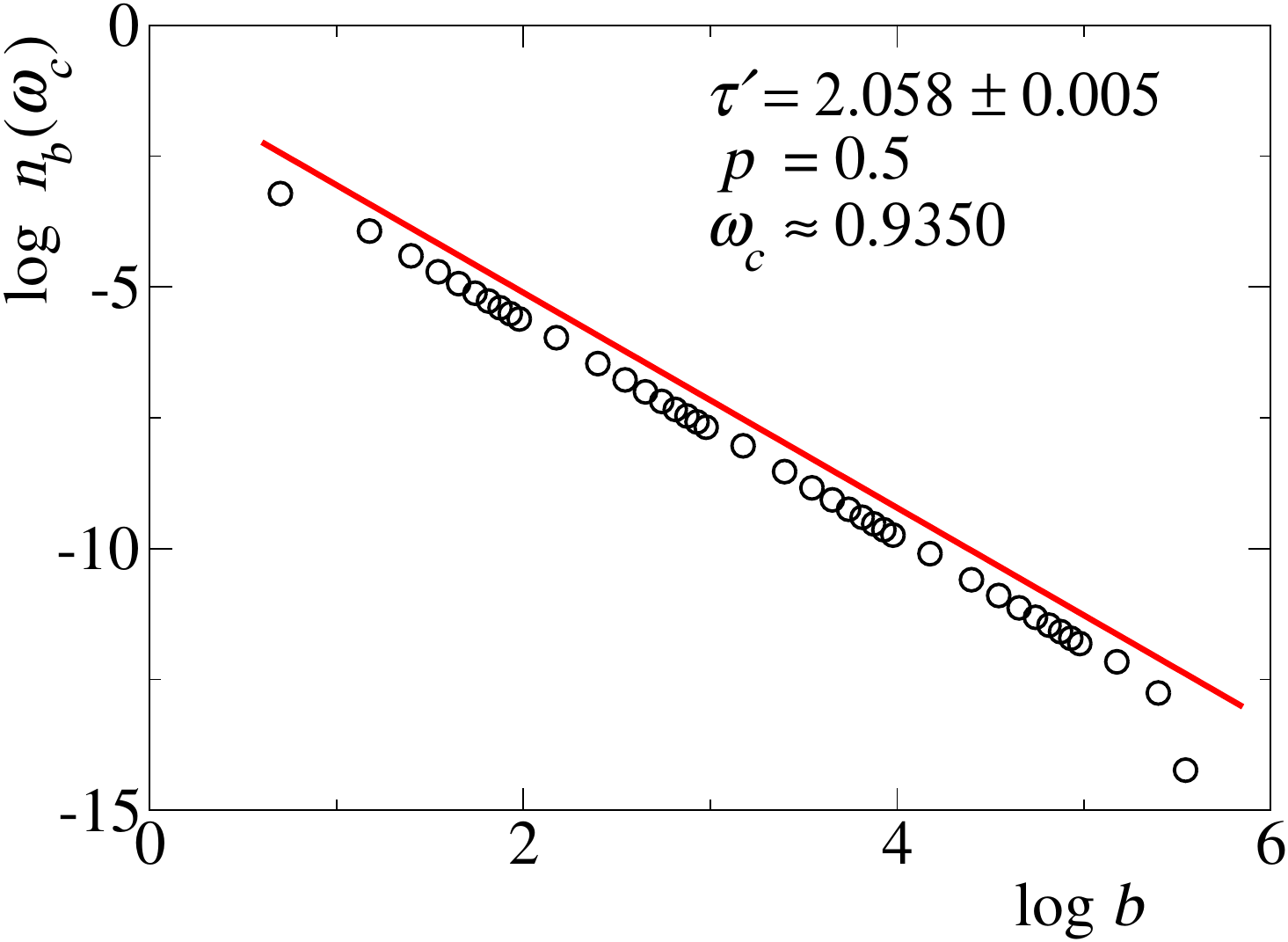}\hfill
    \includegraphics[width=0.33\linewidth,clip]{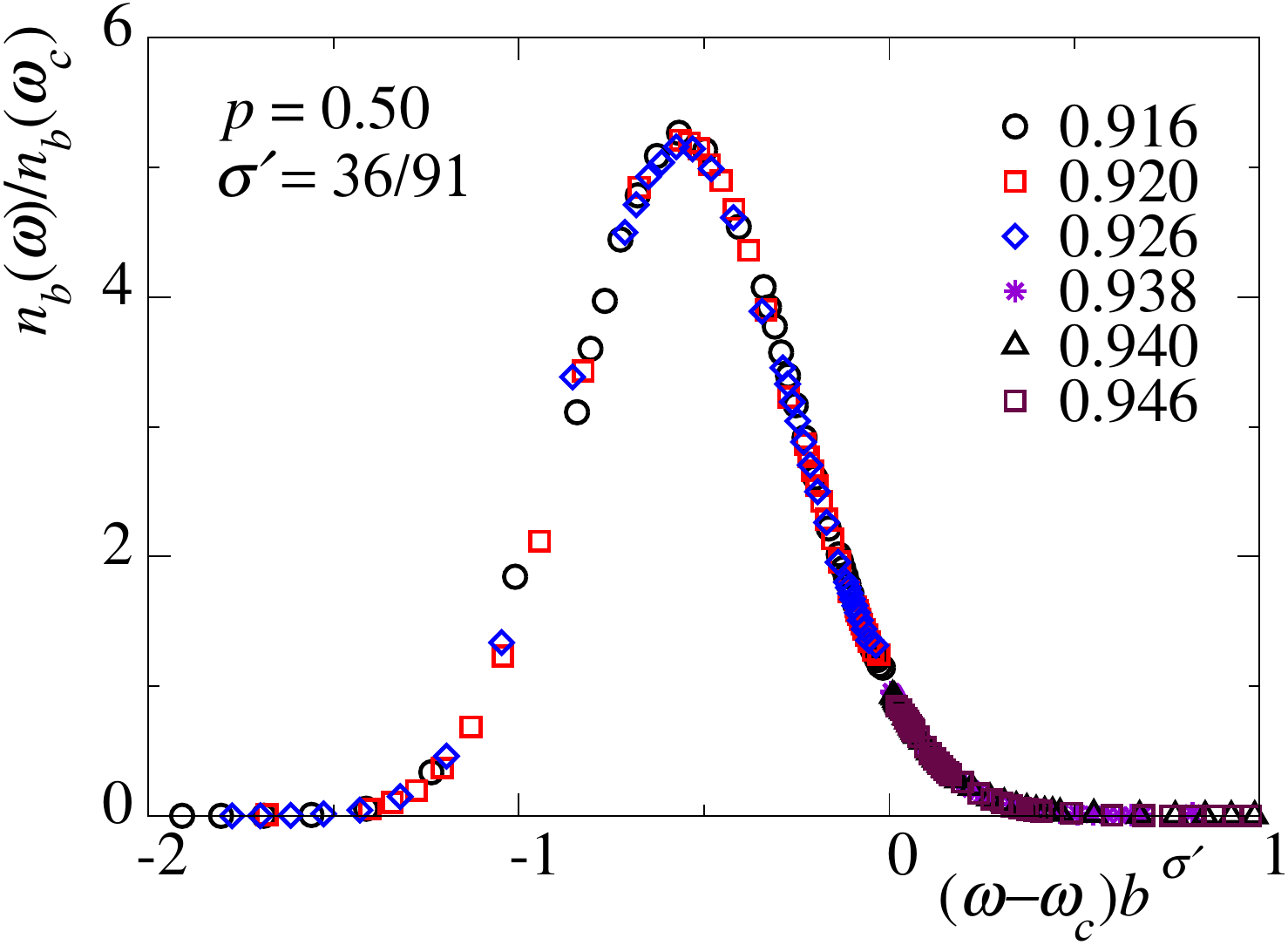}\hfill \includegraphics[width=0.33\linewidth,clip]{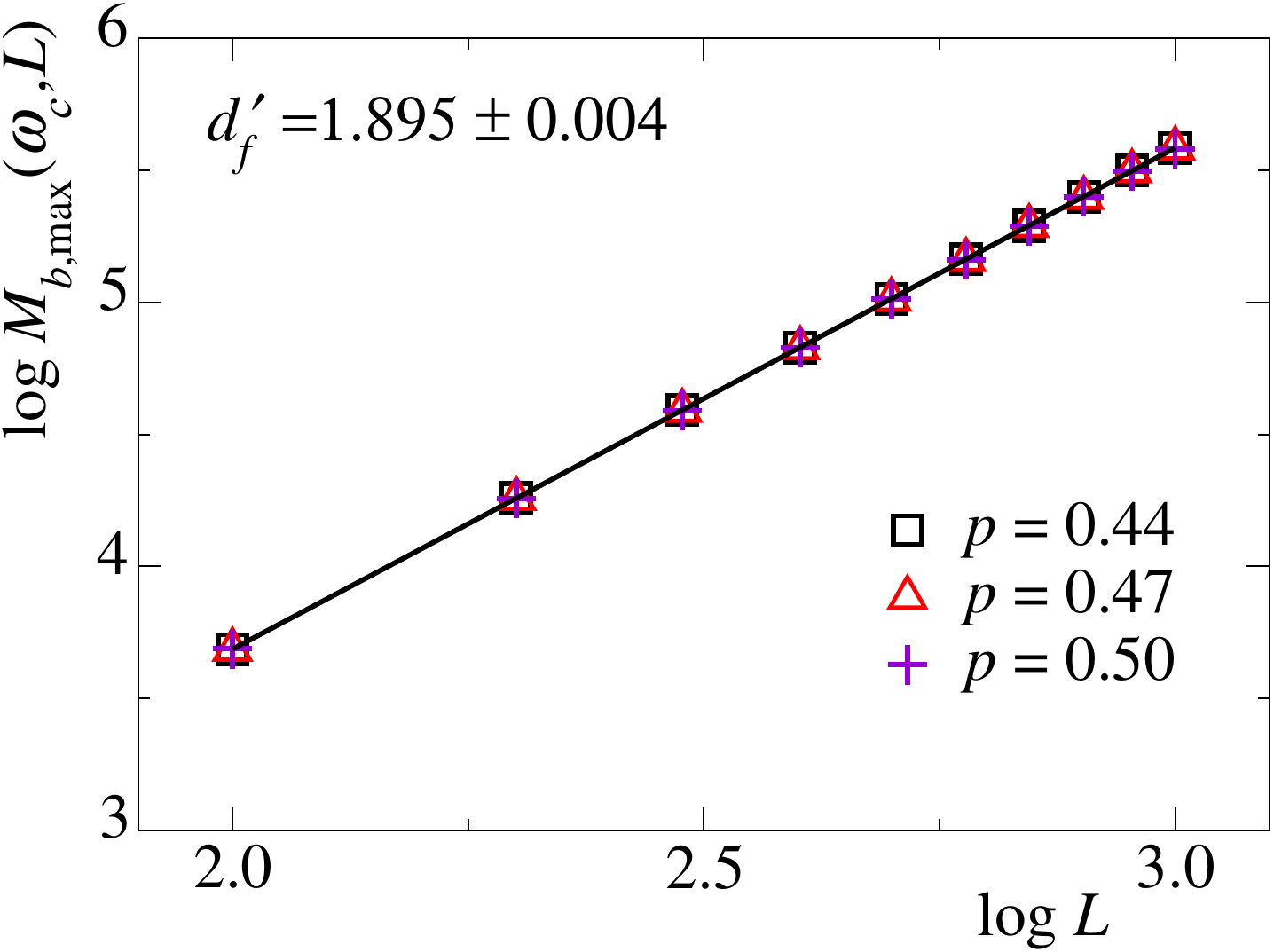}\hfill}
     \centerline{\hfill (a) \hfill \hfill (b) \hfill\hfill (c) \hfill}
    \caption{\label{spcls} (a) Plot of cluster size distribution $n_b(\omega_c)$ against the size of the clusters $b$ at the threshold $\omega_c=0.9350$ for $p=0.5$ on a double logarithmic scale for a lattice size $L = 1000$. The red line of slope $91/48$ is a guide to the eye. (b) Plot of the scaled cluster size distribution $n_b(\omega)/n_b(\omega_c)$ against the scaled variable $(\omega-\omega_c)b^\sigma$ for $p=0.5$ on a lattice of size $L = 2000$. A good collapse of data is obtained for $\sigma'=\sigma=36/91$. (c) Plot of mass of the spanning cluster $M_{b,\rm max}$  against lattice size $L$ on a double logarithmic scale. The black straight line is the line of linear regression with slope $d_f = 1.895 \pm 0.004$.  }
\end{figure}

Following Eq. \ref{eq-3-19}, the cluster size distribution in RBMWP is assumed to be
\begin{equation}
    \displaystyle   
    n_b(\omega)=b^{-\tau'}\mathcal{N}_b\left[(\omega-\omega_c)b^{\sigma'}\right] \;,
\end{equation}
where $\tau'$, $\sigma'$ are two exponents and $\mathcal{N}_b$ is the scaling function. The cluster distribution at the critical threshold can be written as
\begin{equation}
    \label{eq-5-3}
    \displaystyle
    n_{b}(\omega_{c}) \sim b^{-\tau'} \;,
\end{equation}
where $b$ is the size of the clusters and $\tau'$ is the cluster size distribution exponent. In Fig. \ref{spcls} (a), $n_b(\omega_c)$, for $p=0.5$ and $\omega_c=0.9350$, is plotted against the cluster size $b$ on a double logarithmic scale. The exponent $\tau'$ measured by linear regression of the data points is found to be $\tau'=2.058\pm 0.005$. It is very close to the exponent $\tau=187/91$ as that of ordinary site percolation, shown by a red line as a guide to the eye. It seems $\tau'$ of RBMWP is the same as that $\tau$ of ordinary site percolation, as well as mixed wet percolation. 

Now, we verify the exponent $\sigma'$. The scaled distribution $(n_b(\omega)/n_b(\omega_c)$ is given by
\begin{equation}
    \displaystyle
    \frac{n_b(\omega)}{n_b(\omega_c)}\sim \mathcal{N}_b\left[(\omega-\omega_c)b^{\sigma'}\right] \;,
\end{equation}
where $(\omega-\omega_c)b^{\sigma'}$ is the scaled variable. In Fig. \ref{spcls} (b), the scaled distribution $n_b(\omega)/n_b(\omega_c)$ is plotted against the scaled variable $(\omega-\omega_c)b^{\sigma'}$.  Since $\tau'=\tau=91/48$, we presume that $\sigma'=\sigma=36/91$ is that of ordinary site percolation and mixed wet percolation. It can be seen that a good collapse of data is obtained, suggesting that $\sigma'=\sigma$. 

Although $\nu'$ and $\tau'$ were very close to corresponding exponents of ordinary site percolation, we will verify the fractal dimension of the spanning clusters of RBMWP. A typical spanning cluster of RBMWP is already shown in Fig. \ref{morphology_random} in red colour. Although this spanning cluster has dangling ends, it looks similar to the spanning cluster of Fig. \ref{clusters} (a) shown in red. The mass of the spanning cluster of RBMWP at $\omega = \omega_c$ is supposed to scale with the lattice size $L$ as $M_{b,\rm max}(\omega_c,L) \sim L^{d'_f}$. In Fig. \ref{spcls}(c), we plot the average mass $M_{b,\rm max}$ of the spanning cluster against $L$ in a double logarithmic scale for $p = 0.44,0.47 \text{ and } 0.50$. The black straight line is the line of linear regression with slope $d'_f = 1.895 \pm 0.004 $. It is observed that the average mass of the spanning cluster $M_{b,\rm max}$ at $\omega = \omega_c$ is very close to each other for all values of $p$. They become indistinguishable on a log scale. Though $\omega_c$s are different for different $p$, as the system becomes critical for each $p$, not only does the fractal dimension become the same as that of ordinary site percolation, but also the mass of the spanning cluster becomes the same for all $p$. Thus, the fractal dimension $d'_f=d_f$ remains unchanged. 

Since the exponents $\nu'$, $\tau'$ and the fractal dimension $d'_f$ are similar to those of the ordinary site percolation, RBMWP seems to belong to the same universality class of ordinary site percolation as it is observed for mixed wet percolation. Though these models have different cluster size distributions \(n_b(p)\) and critical masses $B_{\mathrm{max}}(p_c, L)$ of the infinite clusters, their scaling behaviour remains the same with their respective cluster size $b$ or with the system size $L$. That is why all models are in the same universality class.

\subsection{Finite size scaling}
\begin{figure}[t] 
    \centerline{\hfill
        \includegraphics[width=0.33\linewidth,clip]{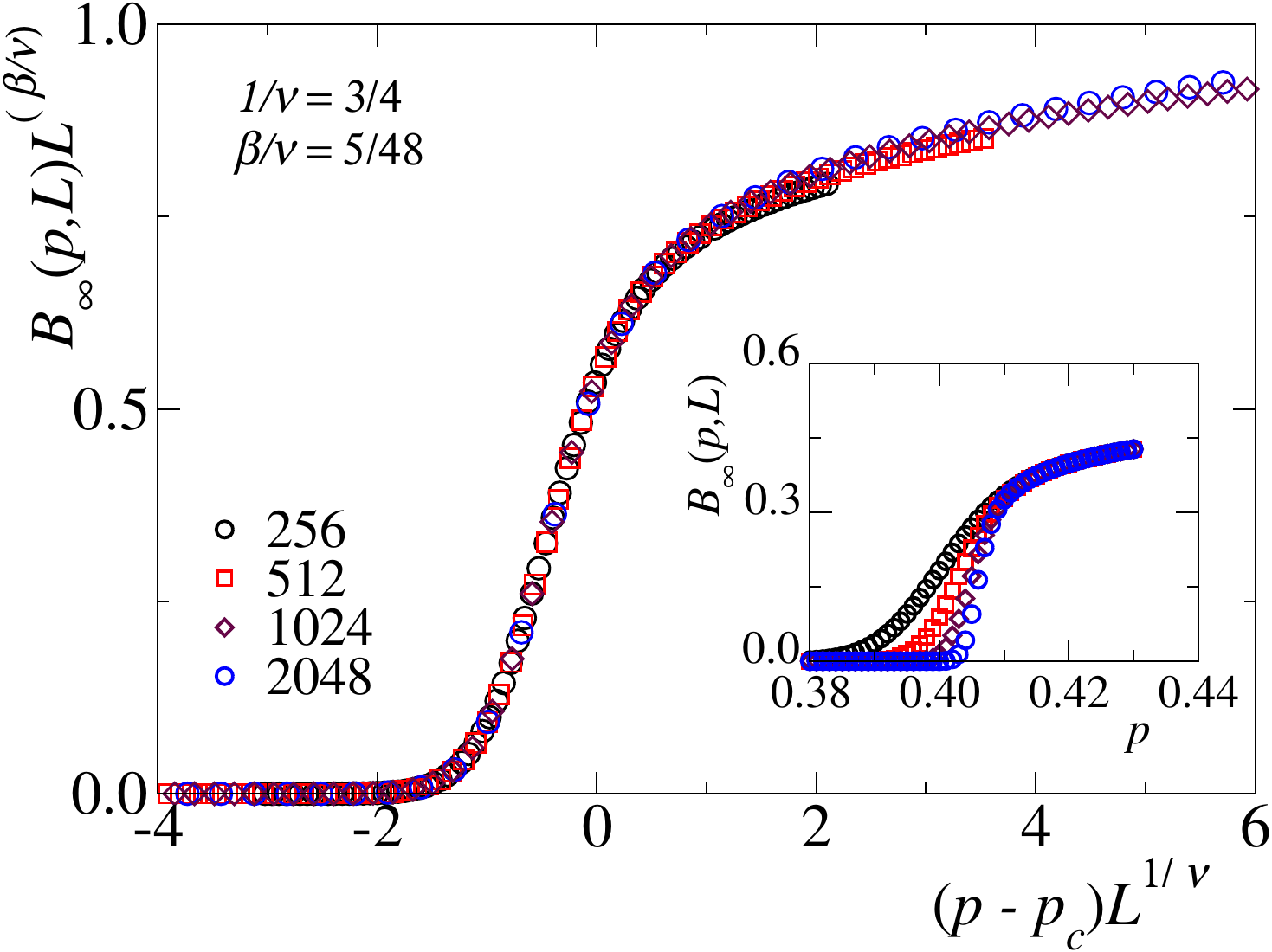}\hfill
        \includegraphics[width=0.33\linewidth,clip]{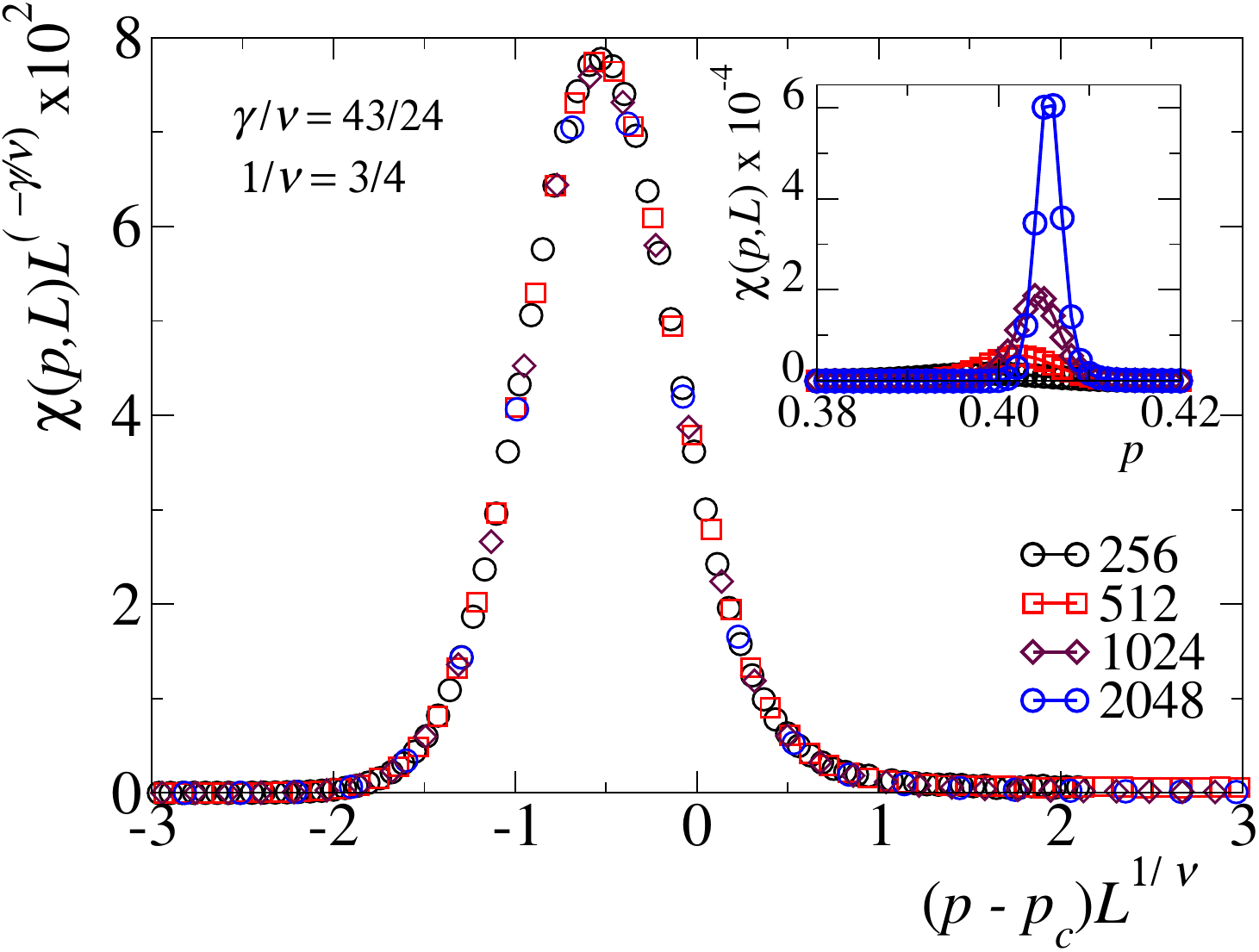}\hfill
        \includegraphics[width=0.33\linewidth,clip]{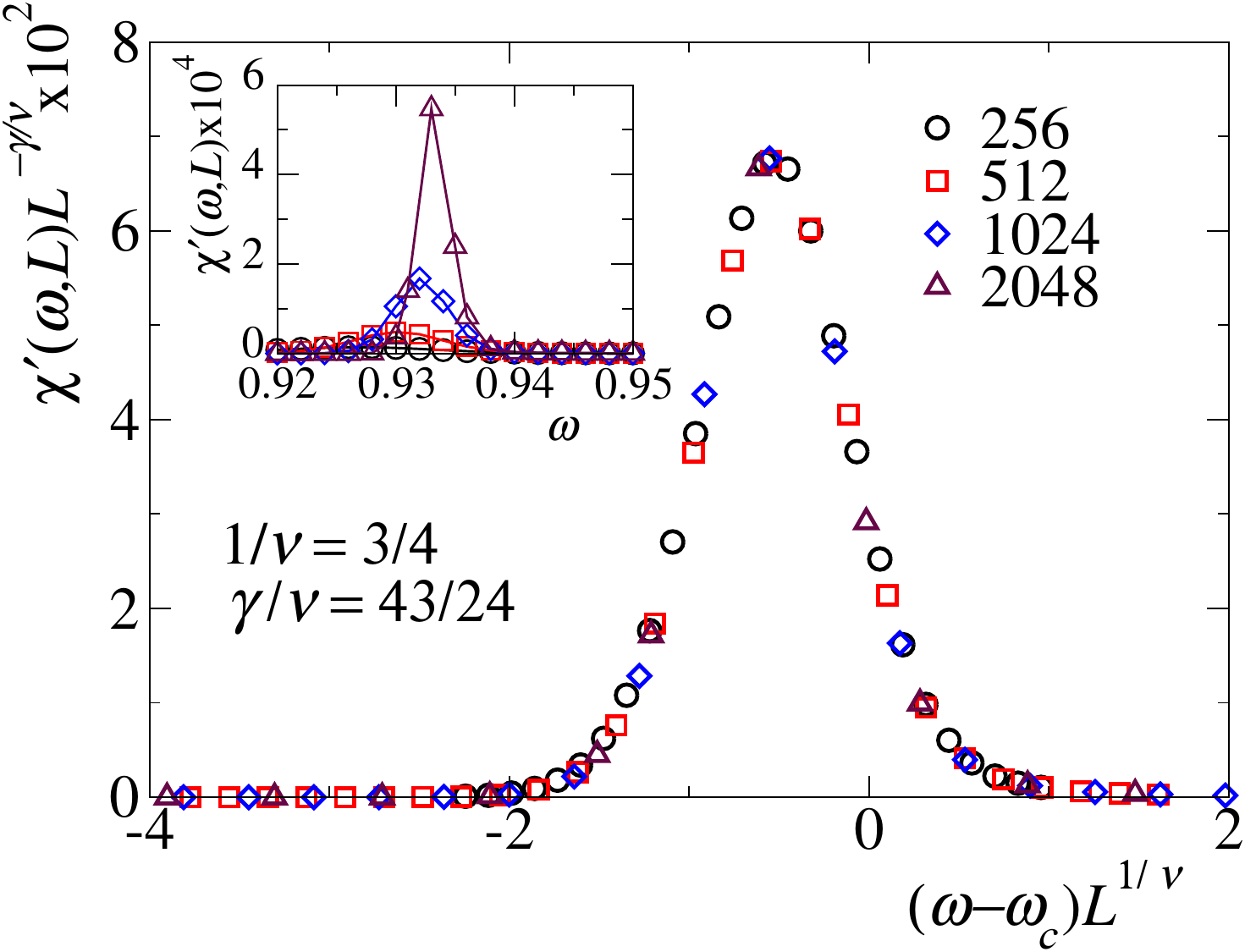}\hfill}
        \centerline{\hfill (a) \hfill \hfill (b) \hfill\hfill (c) \hfill}
    \caption{\label{fssop} (a) Plot of scaled order parameter $B_\infty(p,L)L^{\beta/\nu}$ against the scaled variable $(p-p_c)L^{1/\nu}$ taking $\beta/\nu=5/48$ and $1/\nu=3/4$. Plots of $B_\infty(p,L)$ against $p$ for different system sizes $L$ are in the inset. A good collapse of data is observed with the chosen values of the critical exponents. (b) Plot of scaled fluctuation $\chi(p,L)L^{-\gamma/\nu}$ against the scaled variable $(p-p_c)L^{1/\nu}$ taking $\gamma/\nu=43/24$ and $1/\nu=3/4$. In the inset, plots of $\chi(p,L)$ against $p$ for different system sizes $L$ are shown. A good collapse of data is observed, which indicates the right choice of the critical exponent's values. (c) Plot of scaled fluctuation $\chi'(\omega,L)L^{-\gamma'/\nu'}$ against the scaled variable $(\omega-\omega_c)L^{1/\nu'}$ taking $\gamma'/\nu'=43/24$ and $1/\nu'=3/4$. Plots of $\chi(\omega,L)$ against $\omega$ for different system sizes $L$ are shown in the inset. With the chosen values of the critical exponents, a good collapse of data is observed for the scaled fluctuation against the scaled variable.}
\end{figure}

Now, we test the finite-size scaling of the geometrical quantities. First, we consider the order parameter $B_\infty(p,L) = L^{-\beta/\nu}\widetilde{B}_{\infty}\left[(p-p_c)L^{1/\nu}\right]$ as given in Eq. \ref{eq-4-8}. In Fig. \ref{fssop} (a), we plot the scaled order parameter $B_\infty(p,L)L^{\beta/\nu}$ against the scaled variable $(p-p_c)L^{1/\nu}$ taking $\beta/\nu=5/48$ and $1/\nu=3/4$ as that of ordinary site percolation . The variations of $B_\infty$ against $p$ for four different system sizes $L=256$, $512$, $1024$ and $2048$ are shown in the inset. A good collapse of data of all four system sizes onto a single curve is found for the scaled order parameter $B_\infty(p,L)L^{\beta/\nu}$ against the scaled variable $(p-p_c)L^{1/\nu}$. Thus, the finite size scaling form assumed for $B_\infty(p,L)$ is found to be valid with the right values of the critical exponents $\beta$ and $\nu$.

Next, we study the FSS of the fluctuation $\chi$ in the order parameter. The scaling form of the fluctuation is  $\chi(p,L)=L^{\gamma/\nu}\widetilde{\chi}\left[(p-p_c)L^{1/\nu}\right]$ as given in Eq. \ref{eq-4-9}. In Fig. \ref{fssop} (b), we plot the scaled fluctuation $\chi(p,L)L^{\gamma/\nu}$ against the scaled variable $(p-p_c)L^{1/\nu}$ taking $\gamma/\nu=43/24$ and $1/\nu=3/4$. The plots of $\chi(p,L)$ against $p$ for four different system sizes $L=256$, $512$, $1024$ and $2048$ are shown in the inset. It can be seen that the fluctuation around $p_c$ is diverging as $L$ increases. The scaled fluctuations against the scaled variable are found to have a good collapse of data onto a single curve for all four system sizes. Thus, the finite-size scaling form assumed for $\chi(p,L)$ is found to be valid. Hence, the finite size scaling exponent $\gamma/\nu$ that describes the scaling of $\chi(p,L)$ with $L$ is the same as that of the ordinary percolation. The scaling relation $2\beta/\nu+\gamma/\nu=d$ and all other scaling relations involving $\gamma$ and $\nu$ are automatically valid. 

Finally, we verify the FSS form of the fluctuation $\chi'$ in the order parameter of RBMWP. The scaling form of the fluctuation is  $\chi'(\omega,L)=L^{\gamma'/\nu'}\widetilde{\chi'}\left[(\omega-\omega_c)L^{1/\nu'}\right]$. We plot The scaled fluctuation $\chi'(\omega,L)L^{\gamma'/\nu'}$ is plotted against the scaled variable $(\omega-\omega_c)L^{1/\nu'}$ in Fig. \ref{fssop}(c) taking $\gamma'/\nu'=43/24$ as that of ordinary site percolation . It is already found that $1/\nu'=1/\nu=3/4$. The variation of the fluctuation $\chi'(\omega,L)$ against $\omega$ for different $L$ ($256$, $512$, $1024$ and $2048$) is shown in the inset. We found a good collapse of data onto a single curve for all four system sizes. Thus, the finite-size scaling form assumed for $\chi'(\omega,L)$ is found to be valid. Thus, the finite size scaling exponent $\gamma'/\nu'$ that describes the scaling of $\chi(\omega,L)$ with $L$ is the same as that of the ordinary site percolation and mixed wet percolation. Hence, the scaling relations involving $\gamma$', $\nu'$ and other exponents will be automatically valid. 

It is surprising that mixed wet percolation with a completely different cluster geometry than that of bond percolation exhibits critical behaviour with exactly the same values of the critical exponents as well as the fractal dimension. The non-linear scaling given in Eq. \ref{sc1} ($\phi(\rho)\sim\left|\Delta\rho\right|^{\pm q}$) does not lead to any new critical exponent. Moreover, even after imposing randomness to the bond occupation in the RBMWP, the critical behaviour does not change. Both the models, mixed wet percolation and RBMWP, belong to the same universality class as that of ordinary percolation. Interestingly, the loop percolation in a different way studied by Pfeiffer and Rieger \cite{pfeiffer2003critical} is also found to belong to the ordinary percolation universality class. However, spiral lattice animals \cite{santra1989statistics} and spiral percolation \cite{santrs1992spiral} with loops in either clockwise or anticlockwise directions belong to the spiral percolation universality class, different from ordinary percolation. 

\section{Summary and Conclusion}
\label{conclusion}
We presented a percolation model in the context of two-phase flow in a mixed-wet porous medium, where the porous medium is made of a random mixture of two types of grains with opposite wetting properties. The grains are modeled by ordinary site percolation, whereas the bonds in the dual lattice that appear between two adjacent occupied-unoccupied sites on the primal lattice represent the network of pores with no capillary forces. It has been found that spanning clusters of such zero capillary barrier bonds occur between the two critical thresholds, $p_c = 0.4072$, the next-nearest neighbour site percolation threshold and $1-p_c = 0.5927$, the ordinary site percolation threshold. At these two thresholds, one needs to occupy the bonds on the dual lattice with unit probability to have mixed wet percolation. The bond clusters in the dual lattice are found to be the perimeters of the clusters of either the occupied sites or the unoccupied sites on the primal lattice. Though the perimeter bond clusters consist of a number of cluster hulls, exterior and interior, connected by knots and are very different from an ordinary bond percolation cluster, the critical property remains the same as that of OSP. Looking at the perimeter bond clusters, one would expect the critical behaviour to be that of a percolation hull rather than percolation clusters. Since at these points, the randomness and tortuosity of the bond clusters in the dual lattice come from the randomness and tortuosity of the site clusters on the primal lattice, and hence we achieved an unexpected result. Even after the introduction of randomness in the bond occupation in the region, $p_c<p<(1-p_c)$, the critical behaviour remains the same.  However, we found a continuous line of percolation thresholds for mixed wet percolation between the two extreme critical points defined by $p_c$ and $(1-p_c)$.

\section*{Acknowledgment}
The authors are thankful to the Department of Science and Technology for the computational facilities under Grant No. SR/FST/P-11/020/2009 and Param Kamrupa of IIT Guwahati for the generous allowance of CPU hours. AH thanks the Physics Department of the Indian Institute of Technology Guwahati for the appointment as honorary faculty. This work was partly supported by the Research Council of Norway through its Center of Excellence funding scheme, project number 262644. Further support, also from the Research Council of Norway, was provided through its INTPART program, Project No. 309139. AH and SS acknowledge funding from the European Research Council (Grant Agreement 101141323 AGIPORE).  

\bibliography{references.bib}{}

\begin{thebibliography}{10}
\expandafter\ifx\csname url\endcsname\relax
  \def\url#1{\texttt{#1}}\fi
\expandafter\ifx\csname urlprefix\endcsname\relax\def\urlprefix{URL }\fi
\expandafter\ifx\csname href\endcsname\relax
  \def\href#1#2{#2} \def\path#1{#1}\fi

\bibitem{broadbent1957percolation}
S.~R. Broadbent, J.~M. Hammersley, Percolation processes: I. crystals and
  mazes, Math. Proc. Camb. Philos. Soc. 53 (1957) 629.
\newblock \href {http://dx.doi.org/10.1017/S0305004100032680}
  {\path{doi:10.1017/S0305004100032680}}.

\bibitem{stauffer2018introduction}
D.~Stauffer, A.~Aharony, Introduction to percolation theory, Taylor \& Francis,
  London, 1994.
\newblock \href {http://dx.doi.org/10.1201/9781315274386}
  {\path{doi:10.1201/9781315274386}}.

\bibitem{complexity}
K.~Christensen, N.~R. Moloney, Complexity and criticality, World Scientific
  Publishing Company, London, 2005.
\newblock \href {http://dx.doi.org/10.1142/p365} {\path{doi:10.1142/p365}}.

\bibitem{ambegaokar1971hopping}
V.~Ambegaokar, B.~I. Halperin, J.~S. Langer, Hopping conductivity in disordered
  systems, Phys. Rev. B 4 (1971) 2612.
\newblock \href {http://dx.doi.org/10.1103/PhysRevB.4.2612}
  {\path{doi:10.1103/PhysRevB.4.2612}}.

\bibitem{katz1986quantitative}
A.~J. Katz, A.~H. Thompson, Quantitative prediction of permeability in porous
  rock, Phys. Rev. B 34 (1986) 8179(R).
\newblock \href {http://dx.doi.org/10.1103/PhysRevB.34.8179}
  {\path{doi:10.1103/PhysRevB.34.8179}}.

\bibitem{wilkinson1983invasion}
D.~Wilkinson, J.~F. Willemsen, Invasion percolation: a new form of percolation
  theory, J. Phys. A: Math. Gen. 16 (1983) 3365.
\newblock \href {http://dx.doi.org/10.1088/0305-4470/16/14/028}
  {\path{doi:10.1088/0305-4470/16/14/028}}.

\bibitem{achlioptas2009explosive}
D.~Achlioptas, R.~M. D'Souza, J.~Spencer, Explosive percolation in random
  networks, Science 323 (2009) 1453.
\newblock \href {http://dx.doi.org/10.1126/science.1167782}
  {\path{doi:10.1126/science.1167782}}.

\bibitem{riordan2011explosive}
O.~Riordan, L.~Warnke, Explosive percolation is continuous, Science 333 (2011)
  322.
\newblock \href {http://dx.doi.org/10.1126/science.1206241}
  {\path{doi:10.1126/science.1206241}}.

\bibitem{sahimi2011flow}
M.~Sahimi, Flow and transport in porous media and fractured rock: from
  classical methods to modern approaches, WILEY-VCH Verlag GmbH \& Co. KGaA,
  Weinheim, Germany, 2011.
\newblock \href {http://dx.doi.org/10.1002/9783527636693}
  {\path{doi:10.1002/9783527636693}}.

\bibitem{blunt2017multiphase}
M.~J. Blunt, Multiphase flow in permeable media: A pore-scale perspective,
  Cambridge University Press, Cambridge, 2017.
\newblock \href {http://dx.doi.org/10.1017/9781316145098}
  {\path{doi:10.1017/9781316145098}}.

\bibitem{feder2022physics}
J.~Feder, E.~G. Flekk{\o}y, A.~Hansen, Physics of Flow in Porous Media,
  Cambridge University Press, Cambridge, 2022.
\newblock \href {http://dx.doi.org/10.1017/9781009100717}
  {\path{doi:10.1017/9781009100717}}.

\bibitem{hs17}
A.~G. Hunt, M.~Sahimi, Flow, transport, and reaction in porous media:
  Percolation scaling, critical-path analysis, and effective medium
  approximation, Rev. Geophys. 55 (2017) 993.
\newblock \href {http://dx.doi.org/10.1002/2017RG000558}
  {\path{doi:10.1002/2017RG000558}}.

\bibitem{ga19}
F.~Guo, S.~A. Aryana, An experimental investigation of flow regimes in
  imbibition and drainage using a microfluidic platform, Energies 12 (2019)
  1390.
\newblock \href {http://dx.doi.org/10.3390/en12071390}
  {\path{doi:10.3390/en12071390}}.

\bibitem{ltz88}
R.~Lenormand, E.~Touboul, Zarcone, Numerical models and experiments on
  immiscible displacements in porous media, J. Fluid Mech. 189 (1988) 165.
\newblock \href {http://dx.doi.org/10.1017/S0022112088000953}
  {\path{doi:10.1017/S0022112088000953}}.

\bibitem{lz89}
R.~Lenormand, C.~Zarcone, Capillary fingering: Percolation and fractal
  dimension, Transp. Porous Med. 4 (1989) 599.
\newblock \href {http://dx.doi.org/10.1007/BF00223630}
  {\path{doi:10.1007/BF00223630}}.

\bibitem{w86}
D.~Wilkinson, Percolation effects in immiscible displacement, Phys. Rev. A 34
  (1986) 1380.
\newblock \href {http://dx.doi.org/10.1103/PhysRevA.34.1380}
  {\path{doi:10.1103/PhysRevA.34.1380}}.

\bibitem{irannezhad2023fluid}
A.~Irannezhad, B.~K. Primkulov, R.~Juanes, B.~Zhao, Fluid-fluid displacement in
  mixed-wet porous media, Phys. Rev. Fluids 8 (2023) L012301.
\newblock \href {http://dx.doi.org/10.1103/PhysRevFluids.8.L012301}
  {\path{doi:10.1103/PhysRevFluids.8.L012301}}.

\bibitem{irannezhad2023characteristics}
A.~Irannezhad, B.~K. Primkulov, R.~Juanes, B.~Zhao, Characteristics of
  fluid–fluid displacement in model mixed-wet porous media: patterns,
  pressures and scalings, J. Fluid Mech. 967 (2023) A27.
\newblock \href {http://dx.doi.org/10.1017/jfm.2023.500}
  {\path{doi:10.1017/jfm.2023.500}}.

\bibitem{fyhn2023effective}
H.~Fyhn, S.~Sinha, A.~Hansen, Effective rheology of immiscible two-phase flow
  in porous media consisting of random mixtures of grains having two types of
  wetting properties, Front. Phys. 11 (2023) 1175426.
\newblock \href {http://dx.doi.org/10.3389/fphy.2023.1175426}
  {\path{doi:10.3389/fphy.2023.1175426}}.

\bibitem{geistlinger2021new}
H.~Geistlinger, B.~Zulfiqar, S.~Schlueter, M.~Amro, New structural percolation
  transition in fractional wet 3d-porous media: A comparative $\mu$-ct study,
  Water Resources Research 57 (2021) e2021WR030037.
\newblock \href {http://dx.doi.org/10.1029/2021WR030037}
  {\path{doi:10.1029/2021WR030037}}.

\bibitem{sgv21}
S.~Sinha, M.~A. Gjennestad, M.~Vassvik, A.~Hansen, Fluid meniscus algorithms
  for dynamic pore-network modeling of immiscible two-phase flow in porous
  media, Front. Phys. 9 (2021) 548497.
\newblock \href {http://dx.doi.org/10.3389/fphy.2020.548497}
  {\path{doi:10.3389/fphy.2020.548497}}.

\bibitem{erpelding2013history}
M.~Erpelding, S.~Sinha, K.~T. Tallakstad, A.~Hansen, E.~G. Flekk{\o}y, K.~J.
  M{\aa}l{\o}y, History independence of steady state in simultaneous two-phase
  flow through two-dimensional porous media, Phys. Rev. E 88 (2013) 053004.
\newblock \href {http://dx.doi.org/10.1103/PhysRevE.88.053004}
  {\path{doi:10.1103/PhysRevE.88.053004}}.

\bibitem{tallakstad2009steady}
K.~T. Tallakstad, H.~A. Knudsen, T.~Ramstad, G.~L{\o}voll, K.~J. M{\aa}l{\o}y,
  R.~Toussaint, E.~G. Flekk{\o}y, Steady-state two-phase flow in porous media:
  Statistics and transport properties, Phys. Rev. Lett. 102 (2009) 074502.
\newblock \href {http://dx.doi.org/10.1103/PhysRevLett.102.074502}
  {\path{doi:10.1103/PhysRevLett.102.074502}}.

\bibitem{tallakstad2009steadyb}
K.~T. Tallakstad, G.~L{\o}voll, H.~A. Knudsen, T.~Ramstad, E.~G. Flekk{\o}y,
  K.~J. M{\aa}l{\o}y, Steady-state, simultaneous two-phase flow in porous
  media: An experimental study, Phys. Rev. E. 80 (2009) 036308.
\newblock \href {http://dx.doi.org/10.1103/PhysRevE.80.036308}
  {\path{doi:10.1103/PhysRevE.80.036308}}.

\bibitem{sinha2012effective}
S.~Sinha, A.~Hansen, Effective rheology of immiscible two-phase flow in porous
  media, Europhys. Lett. 99 (2012) 44004.
\newblock \href {http://dx.doi.org/10.1209/0295-5075/99/44004}
  {\path{doi:10.1209/0295-5075/99/44004}}.

\bibitem{wierman1989ab}
J.~C. Wierman, $\text{AB}$ percolation: A brief survey, Banach Center
  Publications 25 (1989) 241.
\newblock \href {http://dx.doi.org/10.4064/-25-1-241-251}
  {\path{doi:10.4064/-25-1-241-251}}.

\bibitem{wu2003ab}
X.-Y. Wu, S.~Y. Popov, On $\text{AB}$ bond percolation on the square lattice
  and ab site percolation on its line graph, Journal of statistical physics 110
  (2003) 443.
\newblock \href {http://dx.doi.org/10.1023/A:1021091316925}
  {\path{doi:10.1023/A:1021091316925}}.

\bibitem{herrmann1990modelization}
H.~J. Herrmann, S.~Roux, Modelization of fracture in disordered systems, in:
  Statistical models for the fracture of disordered media, Elsevier, 1990, p.
  159.
\newblock \href {http://dx.doi.org/10.1016/B978-0-444-88551-7.50016-1}
  {\path{doi:10.1016/B978-0-444-88551-7.50016-1}}.

\bibitem{hans}
L.~B{\"o}ttcher, H.~J. Herrmann, Computational Statistical Physics, Cambridge
  University Press, 2011.
\newblock \href {http://dx.doi.org/10.1017/9781108882316}
  {\path{doi:10.1017/9781108882316}}.

\bibitem{hk}
J.~Hoshen, R.~Kopelman, Percolation and cluster distribution. i. cluster
  multiple labeling technique and critical concentration algorithm, Phys. Rev.
  B 14 (1976) 3438.
\newblock \href {http://dx.doi.org/10.1103/PhysRevB.14.3438}
  {\path{doi:10.1103/PhysRevB.14.3438}}.

\bibitem{de1986multiscaling}
L.~de~Arcangelis, S.~Redner, A.~Coniglio, Multiscaling approach in random
  resistor and random superconducting networks, Phys. Rev. B 34 (1986) 4656.
\newblock \href {http://dx.doi.org/10.1103/PhysRevB.34.4656}
  {\path{doi:10.1103/PhysRevB.34.4656}}.

\bibitem{stanley1977cluster}
H.~E. Stanley, Cluster shapes at the percolation threshold: and effective
  cluster dimensionality and its connection with critical-point exponents, J.
  Phys. A: Math. Gen. 10 (1977) L211.
\newblock \href {http://dx.doi.org/10.1088/0305-4470/10/11/008}
  {\path{doi:10.1088/0305-4470/10/11/008}}.

\bibitem{fan2021statistical}
J.~Fan, J.~Meng, J.~Ludescher, X.~Chen, Y.~Ashkenazy, J.~Kurths, S.~Havlin,
  H.~J. Schellnhuber, Statistical physics approaches to the complex earth
  system, Phys. Rep. 896 (2021) 1.
\newblock \href {http://dx.doi.org/10.1016/j.physrep.2020.09.005}
  {\path{doi:10.1016/j.physrep.2020.09.005}}.

\bibitem{xun2021site}
Z.~Xun, D.~Hao, R.~M. Ziff, Site percolation on square and simple cubic
  lattices with extended neighborhoods and their continuum limit, Phys. Rev. E
  103 (2021) 022126.
\newblock \href {http://dx.doi.org/10.1103/PhysRevE.103.022126}
  {\path{doi:10.1103/PhysRevE.103.022126}}.

\bibitem{mg05}
K.~Malarz, S.~Galam, Square-lattice site percolation at increasing ranges of
  neighbor bonds, Phys. Rev. E 71 (2005) 016125.
\newblock \href {http://dx.doi.org/0.1103/PhysRevE.71.016125}
  {\path{doi:0.1103/PhysRevE.71.016125}}.

\bibitem{nz00}
M.~E.~J. Newman, R.~M. Ziff, Efficient monte carlo algorithm and high-precision
  results for percolation, Phys. Rev. Lett. 85 (2000) 4104.
\newblock \href {http://dx.doi.org/10.1103/PhysRevLett.85.4104}
  {\path{doi:10.1103/PhysRevLett.85.4104}}.

\bibitem{sorensen}
A.~Malthe-S{\o}renssen, Percolation Theory Using Python, Springer Nature, 2024.
\newblock \href {http://dx.doi.org/10.1007/978-3-031-59900-2}
  {\path{doi:10.1007/978-3-031-59900-2}}.

\bibitem{ziff1986test}
R.~M. Ziff, Test of scaling exponents for percolation-cluster perimeters,
  Physical review letters 56 (1986) 545.
\newblock \href {http://dx.doi.org/10.1103/PhysRevLett.56.545}
  {\path{doi:10.1103/PhysRevLett.56.545}}.

\bibitem{pfeiffer2003critical}
F.~O. Pfeiffer, H.~Rieger, Critical properties of loop percolation models with
  optimization constraints, Phys. Rev. E 67 (2003) 056113.
\newblock \href {http://dx.doi.org/10.1103/PhysRevE.67.056113}
  {\path{doi:10.1103/PhysRevE.67.056113}}.

\bibitem{santra1989statistics}
S.~B. Santra, I.~Bose, Statistics of spiral lattice site animals with loops, J.
  Phys. A: Math. Gen 22 (1989) 5043.
\newblock \href {http://dx.doi.org/10.1088/0305-4470/22/22/028}
  {\path{doi:10.1088/0305-4470/22/22/028}}.

\bibitem{santrs1992spiral}
S.~B. Santra, I.~Bose, Spiral site percolation on the square and triangular
  lattices, J. Phys. A: Math. Gen. 25 (1992) 1105.
\newblock \href {http://dx.doi.org/10.1088/0305-4470/25/5/018}
  {\path{doi:10.1088/0305-4470/25/5/018}}.

\end{thebibliography}
\bigskip
\centerline{------------}
\end{document}